%% file: main.tex
\documentclass[11pt]{article}
\input preamble.tex
\input macros.tex

\renewcommand\hl[1]{#1} 

\graphicspath{[./images/]}

\begin{document}

\input{title}

\date{}

\maketitle

\abstract
\input{abstract}

\section{Introduction}
\label{sec:intro}
\input{intro}

\section{Methods}
\label{sec:methods}
\input{methods}

\section{Results}
\label{sec:results}
\input{results}

\section{Discussion}
\label{sec:discussion}
\input{discussion}

\section{Conclusions}
\label{sec:conclusion}
\input{conclusion}

\section*{Acknowledgments}
\label{sec:acknowledgments}
\input{acknowledgments}

\begin{appendices}
\input{appendix}

\end{appendices}
\bibliographystyle{naturemag}
\bibliography{main_ref.bib}


\end{document}

%% file: preamble.tex
\usepackage{nicematrix}

\usepackage{amsmath}
\usepackage{amsfonts}
\usepackage{amssymb}
\usepackage{hyperref}
\usepackage{subfig}
\usepackage{floatrow}

\usepackage{graphicx}
\usepackage{stix}
\usepackage{soul}

\usepackage{float}
\usepackage{multirow}
\usepackage{placeins}
\usepackage{color}

\usepackage{array}
\usepackage{cancel}
\usepackage[margin=0pt,labelsep=space,font=small,labelfont=bf,textfont=it]{caption}
\usepackage[left=1in,right=1in,top=1in,bottom=1in]{geometry}
\usepackage[export]{adjustbox}
\usepackage[nameinlink,noabbrev]{cleveref}
\crefname{equation}{eq.}{eqs.} 
\Crefname{equation}{Eq.}{Eqs.}
\crefname{figure}{fig.}{figs.}
\Crefname{Figure}{Fig.}{Figs.}
\usepackage{booktabs} 
\usepackage[sort,compress]{cite}
\numberwithin{equation}{section}

\usepackage{algorithm}
\usepackage[noend]{algpseudocode}
\usepackage{float}
\floatstyle{plaintop}
\restylefloat{table}

\usepackage{paralist}
\usepackage[title]{appendix}
\usepackage[textsize=tiny]{todonotes}
\usepackage{soul}

%% file: macros.tex
\newcommand{\specialcell}[2][c]{%
  \begin{tabular}[#1]{@{}c@{}}#2\end{tabular}}

\renewcommand{\div}[1]{\nabla_{#1} \cdot}

\newcommand{\grad}[1]{\nabla_{#1}}

\makeatletter
\newcommand*{\rom}[1]{\expandafter\@slowromancap\romannumeral #1@}
\newcommand{\ksiz}{\ensuremath{\boldsymbol{\xi}}}
\newcommand{\piz}{\ensuremath{\boldsymbol{p}}}
\makeatother

\newcommand{\comment}[1]{}

\DeclareMathAlphabet\mathbfcal{OMS}{cmsy}{b}{n}

\newcommand{\RNum}[1]{\uppercase\expandafter{\romannumeral #1\relax}}

%% file: title.tex
\title{GrainGNN: A dynamic graph neural network for predicting 3D grain microstructure \thanks{Notice:  This manuscript has been authored in part by UT-Battelle, LLC, under contract DE-AC05-00OR22725 with the US Department of Energy (DOE). The US government retains and the publisher, by accepting the article for publication, acknowledges that the US government retains a nonexclusive, paid-up, irrevocable, worldwide license to publish or reproduce the published form of this manuscript or allow others to do so, for US government purposes. DOE will provide public access to these results of federally sponsored research in accordance with the DOE Public Access Plan (http://energy.gov/downloads/doe-public-access-plan).}}
\author{%
  Yigong Qin\thanks{Department of Mechanical Engineering, The University of Texas at Austin, Austin, TX 78712, USA; {\tt ygqin@utexas.edu}} \and
  Stephen DeWitt\thanks{Computational Sciences and Engineering Division, Oak Ridge National Laboratory, Oak Ridge, TN 37831, USA; {\tt dewittsj@ornl.gov}}   \and  \hfill
  Balasubramaniam Radhakrishnan\thanks{Computational Sciences and Engineering Division, Oak Ridge National Laboratory, Oak Ridge, TN 37831, USA; {\tt radhakrishnb@ornl.gov}} \and 
   and  George Biros \thanks{Oden Institute, The University of Texas at Austin, Austin, TX 78712, USA; {\tt gbiros@acm.org}} 
}

%% file: abstract.tex
We propose GrainGNN,  a surrogate model for the evolution of polycrystalline grain structure under rapid solidification conditions in metal additive manufacturing. High fidelity simulations of solidification microstructures are typically performed using multicomponent partial differential equations (PDEs) with moving interfaces. The inherent randomness of the PDE initial conditions (grain seeds) necessitates ensemble simulations to predict microstructure statistics, e.g., grain size, aspect ratio, and crystallographic orientation.   Currently such ensemble simulations are prohibitively expensive and surrogates are necessary.

In GrainGNN, we use a dynamic graph to represent interface motion and topological changes due to grain coarsening. We use a reduced representation of the microstructure using hand-crafted features; we combine pattern finding and altering graph algorithms with two neural networks, a classifier (for topological changes) and a regressor (for interface motion). Both networks have an encoder-decoder architecture; the encoder has a multi-layer transformer long-short-term-memory architecture; the decoder is a single layer perceptron.

We evaluate GrainGNN by comparing it to high-fidelity phase field simulations for in-distribution and out-of-distribution grain configurations for solidification under laser power bed fusion conditions.  GrainGNN results in 80\%--90\% pointwise accuracy; and nearly identical distributions of scalar quantities of interest (QoI)  between phase field and GrainGNN simulations compared using Kolmogorov-Smirnov test. GrainGNN's inference speedup (PyTorch on single x86 CPU)  over a high-fidelity phase field simulation (CUDA on a single NVIDIA A100 GPU) is  150$\times$--2000$\times$ for  100-initial grain problem. Further, using GrainGNN, we model the formation of 11,600 grains in 220 seconds on a single CPU core. 

%% file: intro.tex

\begin{figure}[h!]
 \centering
     \includegraphics[width=1\textwidth]{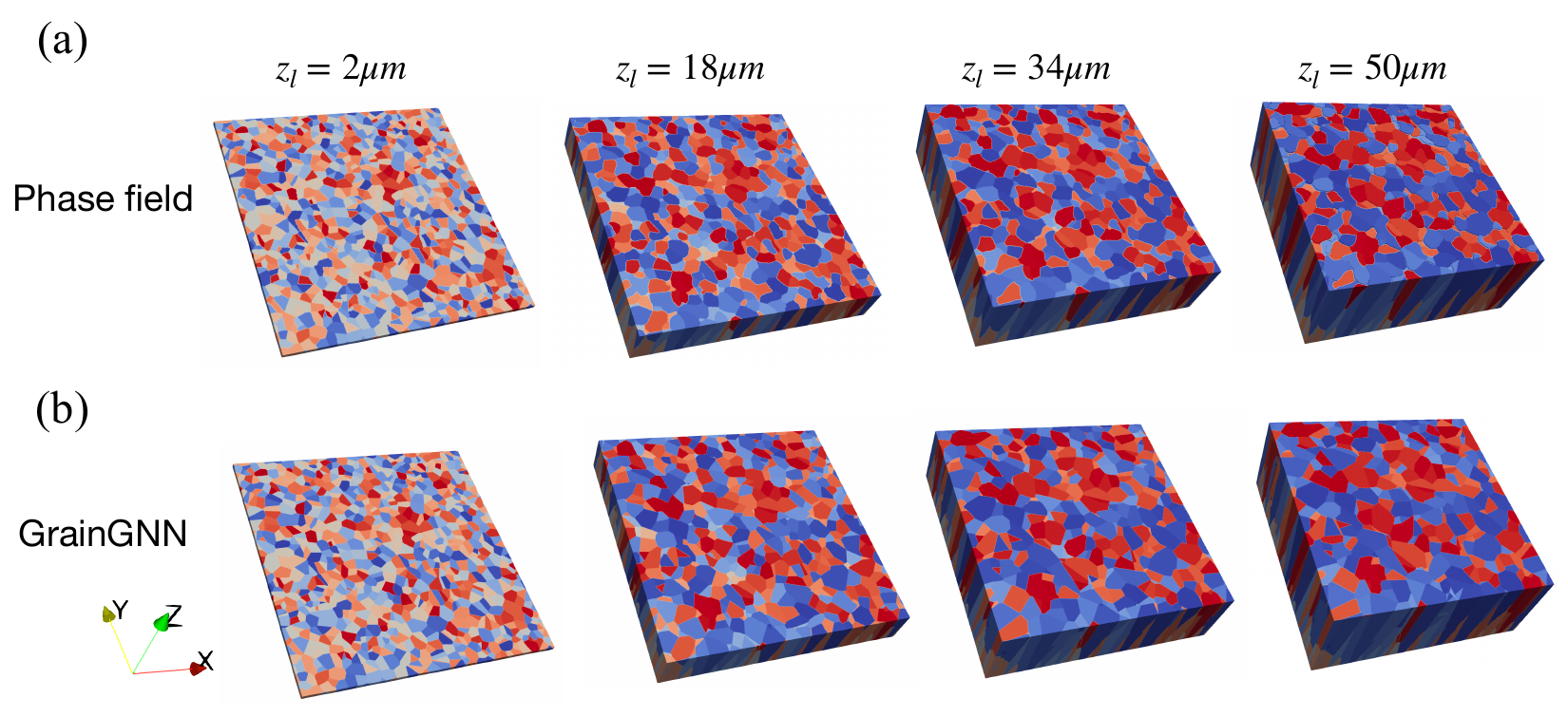}
     \caption{\textbf{Time evolution of the solidified microstructure of stainless steel 316L predicted by a phase field model and GrainGNN.} (a) Phase field simulation. (b) GrainGNN predictions. $z_l$ is the current height of the solid-liquid interface (SLI). The domain size is $(120 \mu m, 120 \mu m, 50\mu m)$. The microstructure of 900 grains at $z_l=2\mu m$ is the input to both phase field simulation and GrainGNN. We apply a temperature field with $G=10K/\mu m$ and $R=2m/s$. The grains are growing in the $z$-direction and stopped at $z_l=50 \mu m$. The elapsed simulation time for the phase field model is 48 minutes on one Nvidia A100 GPU. The GrainGNN inference time is 19 seconds on one Intel x86 6-core CPU. }
     \label{fig:tease}
\end{figure}

Let $\boldsymbol{\theta}(x,y,z,t)$ be the crystal orientation vector as a function of space $(x,y,z)$ and time $t$. We assume we are given a high fidelity model $\mathscr{F}$ such that $\boldsymbol{\theta}(x,y,z,t)=\mathscr{F}(\piz,\ksiz,L)$. Here $\piz$ encodes the meltpool temperature field using two scalars, $G$ and $R$, where $G$ is a scalar metric of the temperature gradient and $R$ is a normalized cooling rate (pooling velocity); $\ksiz$ parameterizes the initial microstructure $\boldsymbol{\theta}(x,y,z,t=0)$ at the solid-liquid interface (SLI); and $L=(L_x,L_y,L_z)$ parameterizes the meltpool size, which we idealize as a 3D rectangular volume with grain epitaxial growth aligned with the  $z$ axis (see \Cref{fig:tease}). In this context, $\boldsymbol{\theta}$ is a piecewise constant function and grains are defined as connected regions with the same $\boldsymbol{\theta}$; $\ksiz$ is a random field, and quantities of interest (QoIs) that depend on $\boldsymbol{\theta}$ can only be statistically characterized and require expectations over $\ksiz$.

Evaluating $\mathscr{F}$ can be extremely expensive as it involves multicomponent, multiscale PDEs with moving interfaces. Furthermore, additive manufacturing (AM) involves a large number of meltpool simulations: the length scale of an AM component can be of the order of meters while the meltpool dimensions are $\mathcal{O}(100\mu$m$)$. \emph{In this context, we seek to find a surrogate model $\widetilde{\boldsymbol{\theta}}=\widetilde{\mathscr{F}}(\piz,\ksiz,L)$ that approximates $\boldsymbol{\theta}$ and the statistics of the quantities of interest. We also require that $\widetilde{\mathscr{F}}$ is much cheaper to compute than $\mathscr{F}$.}

{\bf Background and significance:} High fidelity models are important tools for predicting grain structure formation under rapid solidification conditions existing during AM \cite{smith2016linking}. Common numerical methods include phase field models \cite{steinbach1999generalized,ofori2010quantitative,PINOMAA20201,yang2021phase,CHADWICK2021116862},  cellular automata \cite{gandin19973d,rolchigo2022understanding,rolchigo2022exaca}, and kinetic Monte Carlo methods\cite{rodgers2017simulation} . However, these methods are computationally expensive as they require fine spatial and temporal discretizations. Furthermore, due to the stochastic nature of AM process, ensemble simulations ~\cite{miyoshi2017ultra,miyoshi2021large,chang2017effect,CHADWICK2021116862} are required to statistically characterize quantities of interest (QoIs) like grain size, aspect ratio, and misorientation~\cite{burke1952recrystallization,hillert1965theory,holm2001misorientation}. Tasks such as process control and optimization require repeated invocation of high fidelity models under variable heating source conditions, feedstock composition, and component geometry. These high computational costs 
have restricted the adoption of high fidelity simulations for predicting or controlling the grain structure during AM processes.

{\bf Summary of proposed methodology and contributions:}
Here, following our work in 2D surrogates~\cite{qin2023grainnn}, we focus on epitaxial grain growth from the substrate and ignoring grain nucleation and the evolution of the solute concentration field that contributes to intragranular features of the solidification microstructure including primary and secondary arms, microsegregation, etc. 
The key phenomena that need to be captured at this modeling level are grain envelope evolution and grain coarsening through competitive growth. The latter can be decomposed into grain face elimination and grain elimination~\cite{zollner2017topological, torres2015numerical}. The dynamics of $\mathscr{F}$ are quite challenging to capture due to the evolving interface and the topological changes due to coarsening.

In GrainGNN (\Cref{sec:grainGNN}), first we reduce the representation of $\boldsymbol{\theta}(x,y,z,t)$ using aggressive coarsening in space and time.
\begin{inparaenum}[(i)]
\item
We introduce a reduced graph representation of
$\boldsymbol{\theta}$ that captures grain boundaries, size, edge length, and other geometric features per grain (\Cref{sec:2graph}).
\item
Following \cite{zollner2017topological},  we introduce a categorization of topological events that need to be tracked (\Cref{sec:3graph_dynamics}).
\item
We use two neural networks to model their dynamics:  a regression network to capture the interface motion; and a classification network to estimate probabilities of topology-changing events (\Cref{sec:4arch}).
\item
We introduce graph algorithms that combine the network outputs to evolve the microstructure graph topology (\Cref{sec:5graph_update}).
\item 
We introduce an algorithm that reconstructs $\widetilde{\boldsymbol{\theta}}(x,y,z,t)$ from the reduced graph/feature-based representation of the microstructure (\Cref{sec:6graphTOimage}).
\end{inparaenum}

GrainGNN is calibrated (trained) using pairs $\{\boldsymbol{\theta}_k(x,y,z_{l-1},t_{l-1}),\boldsymbol{\theta}_k(x,y,z_l,t_l)\}_{k}$, i.e., cross-sections of the microstructure at SLI heights $z_{l-1}$ and $z_l$ observed at times $t_{l-1}$ and $t_l$ with $t_l>t_{l-1}$; and $\boldsymbol{\theta}_k(x,y,z,t)=\mathscr{F}(\piz_k,\ksiz_k,L_k)$, where $k$ indicates sampling of AM conditions and initial grain microstructure. We used  $k\approx$40K obtained from $\sim$1,500 high fidelity simulations. 

Using these steps, GrainGNN (\Cref{alg:graingnn}) predicts 3D grain microstructure formation under additive manufacturing conditions. Given the grain orientation at $t=0,z=0$ with a liquid region for $z>0$  (see \Cref{fig:1phase_field}a) and the prediction is $\widetilde{\boldsymbol{\theta}}(x,y,z,t)$ (see \Cref{fig:1phase_field}b). The error is measured by pointwise mismatch and quantities of interest that include the percentage of eliminated grains, grain size distribution, and volume-averaged misorientation.

Our contributions can be summarized as follows.
\begin{itemize}
    \item We introduce a graph and hand-crafted features that greatly compress the spatial representations of the grain microstructure. We create $\boldsymbol{\theta}$-to-graph and graph-to-$\boldsymbol{\theta}$ mappings.
    \item We introduce two graph-transformer long-short-term-memory (LSTM) networks to predict graph feature evolution and grain topological events.
    \item We propose a graph update algorithm with $O(\#\text{grains})$ complexity to reconstruct the next graph with the output of GrainGNN. We generalize the algorithm to predict grain microstructure with unseen $G$, $R$, domain size, grain size distribution, and a larger number of grains.
\end{itemize}

Ground truth data is generated using an established phase-field-based model of epitaxial grain growth (\Cref{sec:pf}). We remark that GrainGNN is independent of the underlying solver/formulation. We train GrainGNN assuming a probability density distribution for $\ksiz$, a fixed $L=L^0$, and a grid-sampled $\piz$ (\Cref{sec:training}). Then we evaluate the \emph{in-distribution} generalization of GrainGNN in which we keep $L=L^0$, sample training-unseen values of $\ksiz$ from the training distribution, and unseen values of $\piz$ and measure pointwise errors of $\boldsymbol{\theta}-\boldsymbol{\widetilde{\theta}}$ as well errors in the statistics of QoIs (\Cref{sec:indis}). We also test the \emph{out-of-distribution} generalization GrainGNN with $L\neq L^0$ and $\ksiz$ sampled from a different distribution (\Cref{sec:outofdis}). We discuss the results, extensions, and limitations of GrainGNN in \Cref{sec:discussion}.

{\bf Related work:}
Surrogate models provide an alternative to high fidelity simulations, with the potential for reduced computational cost. Trained on either high fidelity simulation results or experimental results, such surrogates have been successful in addressing challenges in uncertainty quantification and optimization for a range of applications  \cite{ohayon2014advanced,frangos2010surrogate,legresley2000airfoil,noack2011reduced}. Specifically in the area of material microstructure prediction, machine learning surrogates have received increasing attention in the past few years. Convolutional neural networks have been used for 2D and 3D microstructure image reconstruction \cite{li2018transfer,bostanabad2020reconstruction}. Generative adversarial networks \cite{goodfellow2014generative} are capable of learning microstructure statistics, and can approximate structure-to-material properties forward and inverse maps \cite{yang2018microstructural,lee2021fast}. Recurrent neural networks (RNNs) or its subclass long short-term memory (LSTM) networks \cite{hochreiter1997long} have been used to rapidly predict the time evolution of the microstructure. Existing works consider a representative volume element and use \hl{LSTM} networks to capture evolution statistics for use cases such as: spinoidal decomposition of binary mixtures~\cite{Montes-de-Oca-Zapiain:2021uj,hu2022accelerating}, brittle fracture \cite{yang2021self}, single dendrite growth~\cite{yang2021self}, and grain formation \cite{qin2023grainnn}. Ref. \cite{oommen2022learning} applied the convolutional autoencoder to map the microstructure of a two phase mixture to a latent space and used DeepONet \cite{lu2021learning} to learn the evolution.

Our previous work \cite{qin2023grainnn} introduced GrainNN, a transformer-based LSTM to predict 2D epitaxial grain growth during solidification. GrainNN reduces the computational cost by tracking only the grain boundaries instead of each grid point inside the grains. It evolves manually crafted grain shape descriptors defined on each grain. GrainNN can predict microstructure with low pointwise error while achieving significant speed up over high fidelity simulations. It also utilizes domain decomposition and rectangular-to-curvilinear domain mappings to handle systems of many grains and circular geometry. However, in this 2D setting the grain coupling is only one-dimensional and perpendicular to the temperature gradient, which makes GrainNN hard to model the substantially more complex grain-grain interactions in 3D. 

To generalize GrainNN to 3D, we utilize graph representations of the grain structure that consist of both grains and the vertices of junctions between grains. Vertex representations have previously been used in models \cite{syha2009generalized, kawasaki1989vertex, wakai2000three, torres2015numerical} to track the motion of the grain junction points in polycrystalline microstructures. In these models, the motion of vertices during grain growth is derived from the minimization of the isotropic or anisotropic grain boundary energy. Although for vertex models the update of a grain network is efficient and schemes for handling topological transitions are proposed \cite{syha2009generalized} for curvature-driven grain coarsening, vertex models cannot currently describe grain evolution driven by a temperature gradient during AM solidification. A grain-centric graph structure has been used to predict grain microstructure evolution during solidification \cite{xue2022physics}. This physics-embedded graph network (PEGN) approach combines classic phase-field (PF) theory into a graph representation of the grains to accelerate PF simulations. Evolution is dictated by the minimization of a PF-derived free energy functional. PEGN is able to capture grain statistics in various AM setups but loses the accuracy of PF in terms of predicting actual grain shapes, due to the lack of tracking grain boundaries and topological changes. 

Graph representations of grains have been used for graph neural networks (GNNs) \cite{dai2021graph, hestroffer2023graph} to create static microstructure-to-material-property maps. Graph convolutions on the grain networks are used to capture the spatial variations of microstructure properties and the graph sparsity enables faster evaluations of material behavior than high fidelity models. Beyond applications to material microstructures, GNNs have been successful in a wide range of other applications where the data is amenable to a graph structure \cite{kipf2016semi,scarselli2008graph,morris2019weisfeiler,wieder2020compact,duvenaud2015convolutional,gilmer2017neural,xie2018crystal}.

Previous GNNs successful for processing dynamic graphs include
GC-LSTM \cite{chen2022gc}, EvolveGCN \cite{pareja2020evolvegcn}, and dyngraph \cite{goyal2020dyngraph2vec}, etc. Most use an RNN (LSTM) or autoencoder to encode graph features and a decoder to predict the probability of a node/link formation or destruction. In this paper, we add a transformer \cite{vaswani2017attention} operator to GC-LSTM encoder to capture grain interactions more accurately.


%% file: methods.tex
\begin{figure}[h]
 \centering
     \includegraphics[width=1\textwidth]{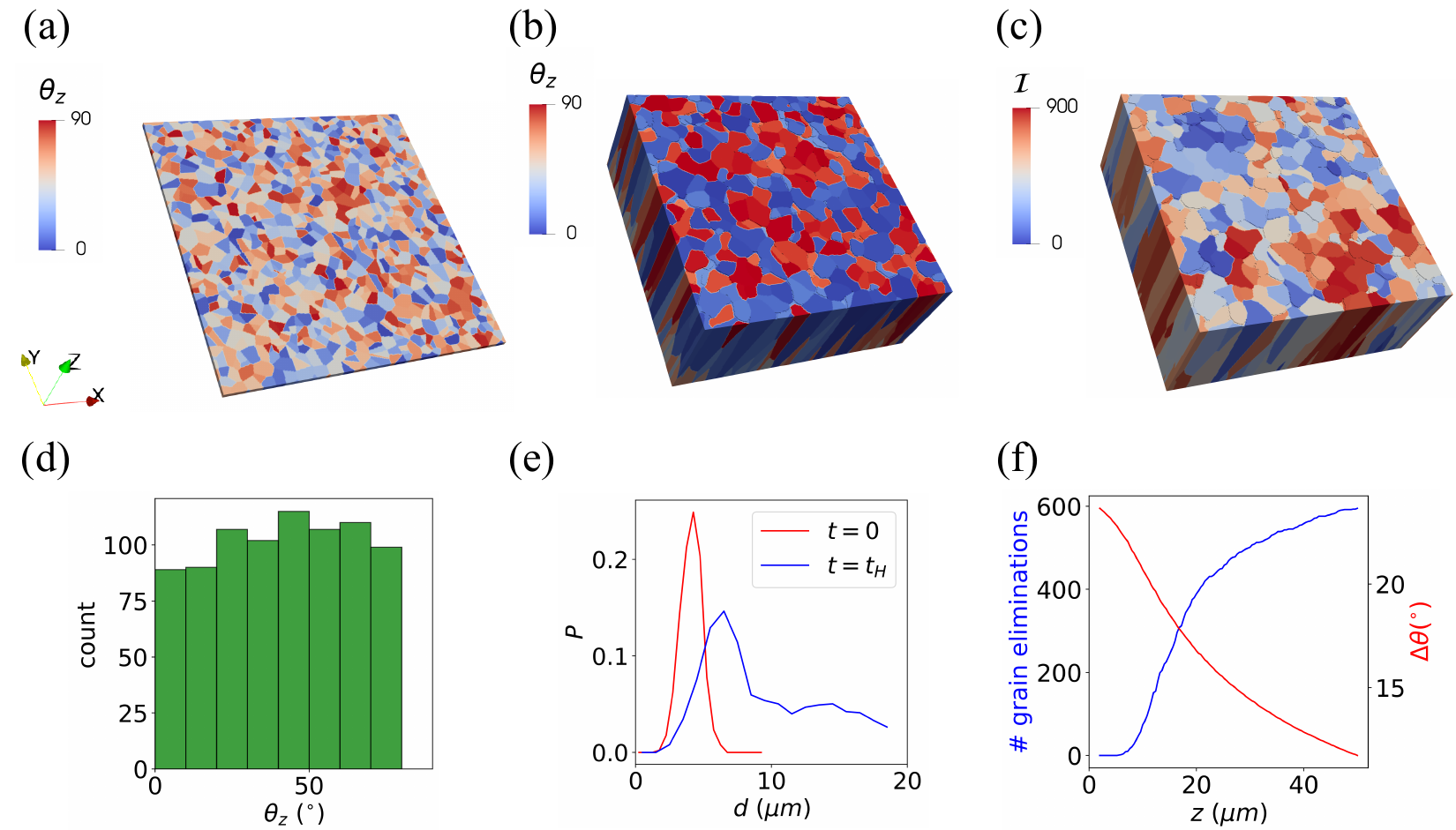}
     \caption{\textbf{A phase field simulation of epitaxial growth of 900 grains.} Temperature gradient $G_z=10\ K/\mu m$ and pulling velocity $R_z=2\ m/s$ (a) Initial substrate with width $120 \mu m$ and height $2\mu m$. $\boldsymbol{\theta}_z$ is the angle between the crystal orientation and the $z$-axis. (b) Phase field microstructure after 10K time steps. The time step size is 2.41 nanoseconds. The final height of the interface is 50 $\mu m$. (c) The corresponding grain index field of (b). Each grain is assigned a unique index from 1 to 900. (d) Initial grain orientation distribution. (e) Probability distributions of grain size $d$ at $t=0$ and $t=t_H$. (f) Time evolution of the quantities of interest. $z$ represents the height of the current solid-liquid interface. As the height of the interface increases, the number of eliminated grains increases and the volume-weight misorientation $\Delta \boldsymbol{\theta}$ decreases. }
     \label{fig:1phase_field}
\end{figure}


\begin{table}
\centering
\small
\caption{Notation table.}
\begin{tabular}{|cc|}
\hline
Symbol & Description (Units)\\
\hline
$T$ & temperature field  \\
$\phi_\alpha$ & phase field\\
$\mathcal{I}$ & grain index field \\
$\boldsymbol{\theta}$ & grain orientation field $(^{\circ})$ \\
$L$ & domain size, $L=\{L_x, L_y, L_z\}$ \\
$G_z$ & temperature gradient $(K/\mu m)$ \\
$R_z$ & pulling velocity  $(m/s)$ \\
$\piz$ & process parameters,  $p=\{G_z, R_z\}$ \\
$\ksiz$ & initial state of the substrate \\
$l$ & layer index \\
$n_l$ & number of layers \\
$I_l$ & grain index image at height $z_l$  \\
$G_l$ & graph extracted from $I_l$ \\
$F_l$ & features of graph $G_l$\\
$g$ & grain index \\
$j$ & junction index \\
$V_g$ & grain vertices of graph $G$ \\
$V_j$ & junction vertices of graph $G$ \\
$n_g$ & number of grains of graph $G$ \\
$n_j$ & number of junctions of graph $G$ \\
$E_{jg}$ & junction-grain edges of graph $G$ \\
$E_{jj}$ & junction-junction edges of graph $G$ \\
$N_j$ & grain neighbors of junction $j$ \\
$N_g$ & junction neighbors of grain $g$ \\
\hline
\end{tabular}\label{tab:noatation}
\end{table}

In this section we discuss the overall methodology. We start with the grain growth model, then we discuss the reduced parameterization of $\boldsymbol{\theta}$, the $\boldsymbol{\theta}$-to-graph map, the graph dynamics, the LSTM neural networks, the graph update algorithm, and the graph-to-$\boldsymbol{\theta}$ map. We conclude with details about the phase field solver. 

\subsection{Problem formulation and high fidelity model}
\label{sec:pf}
Here we describe the model we used to generate the training data. We use a phase field simulation. We would like to emphasize that  GrainGNN does not depend on any particular model or numerical method for generating the grain microstructure. For example level sets, different grain formation models, or cellular automata could be used. The main assumption in our surrogate is does not capture nucleation. 

Regarding the particular grain formation model, we adopt a grain-scale phase field model described in Ref. \cite{pinomaa2019process} and ignore nucleation\footnote{We remark that the model we use neglects the concentration field and only considers the evolution of phase fields with temperature fields in the rapid solidification regime. The grain structure can be modeled with even more accurate and even more expensive dendrite-resolving models~\cite{qin2022dendrite}.}. Each phase field component $\phi_{\alpha}$ is associated with one of $n_g$ different crystalline orientations, where $n_g$ is the number of grains. The dynamics of $\phi_{\alpha}$ is governed by the following equation. 
\begin{subequations}
\begin{align}
 &   \tau_{\alpha}  \frac{\partial \phi_{\alpha} }{\partial t}    =  \div{} (W_{\alpha}^2\grad{} \phi_{\alpha})  +  \sum_{j=x,y,z}  \partial_{j} \left[ |\grad{} \phi_{\alpha}|^2 W_{\alpha} \frac{\partial W_{\alpha}}{\partial (\partial_{j} \phi_{\alpha})}  \right] + \phi_{\alpha} - \phi_{\alpha}^3   - \lambda (1-\phi_{\alpha}^2)^2 \frac{T-T_M}{L_f/C_p} \nonumber \\
&  \phantom{  \tau_{\alpha}  \frac{\partial \phi_{\alpha} }{\partial t}  =  }  - \omega \frac{\phi_{\alpha}+1}{2} \sum_{\beta \neq \alpha}\left(\frac{\phi_{\beta}+1}{2}\right)^{2} \!\!, \quad (L_x, L_y, L^{\prime}_z)\times(0,t_H],  \quad \alpha=1,...,n_g,  \label{eq:micro_jsi}  \\
& \text{BCs: } \ \phi_{\alpha}(x=0) = \phi_{\alpha}(x=L_x) \\
& \phantom{BCs:}  \phi_{\alpha}(y=0) = \phi_{\alpha}(y=L_y) \\
& \phantom{BCs:}  {\grad{}}_z\phi_{\alpha} (z=0,L^{\prime}_z)  =0 \\
& \text{IC: }  \phi_{\alpha}(\boldsymbol{x}, t=0) = 
\begin{cases}
  \tanh{} \left( \frac{z_0 - z}{W_0}\right),  \hspace{0.5cm} \text{if } (x,y) \in \Omega_{\alpha}(\ksiz) \\    
  -1, \hspace{2cm} \text{if } (x,y) \notin \Omega_{\alpha}(\ksiz)
\end{cases}
\end{align}
\end{subequations}
where 
\begin{subequations}
\begin{align}
W_{\alpha} &= W_{0}\left(1 -3 \epsilon_{c} + 4 \epsilon_{c}({\grad{}}_x \phi_{\alpha}^4 + {\grad{}}_y\phi_{\alpha}^4 + {\grad{}}_z\phi_{\alpha}^4 )/| {\grad{}} \phi_{\alpha}|^4\right) , \label{eqn:anis0} \\
\tau_{\alpha} &=\tau_{0}\left(1+3 \epsilon_{k}-4 \epsilon_{k}({\grad{}}_x \phi_{\alpha}^4 + {\grad{}}_y\phi_{\alpha}^4 + {\grad{}}_z\phi_{\alpha}^4 )/| {\grad{}} \phi_{\alpha}|^4\right) , \label{eqn:anis1} \\
T&= T_M+G_z(z-R_z t).
\label{eqn:temp}
\end{align}
\end{subequations}
Here $L_x$, $L_y$, $L^{\prime}_z$ are the domain dimensions for simulation and $t_H$ is the time horizon;  $\tau_0$ is the interface attachment time scale;  $W_0$ is the width of the anisotropic interface; 
$\lambda$ is the thermal coupling constant; $L_f$ is the latent heat; $C_p$ is the heat capacity; 
$T_M$ is the melting temperature; $\omega$ is a scalar interaction parameter that sets the repulsive strength between adjacent grains of different orientations \cite{ofori2010quantitative}. $\epsilon_c$ and $\epsilon_k$ are the capillary and kinetic anisotropy coefficients respectively, which are assumed to have a four-fold symmetry. In \Cref{eqn:temp}, we assume the temperature gradient $G_z$ and the pulling velocity $R_z$ are aligned with the $z$-axis and they are constants during the simulation. Under this temperature profile, grains are growing in the $z$-direction. We name $\piz=\{G_z, R_z\}$ the process parameters. $\piz$ are the main free parameters during AM processes given the alloy with its material parameters.  We specify the values for the material parameters in the appendix. The discretization of \Cref{eq:micro_jsi} is discussed in \Cref{sec:forward}.

We use no-flux boundary conditions at the top and bottom surface ($z=0, z=L^{\prime}_z$) and periodic boundary conditions at the four sides of the domain. We initialize $\phi_{\alpha}$ as follows. Assume that the initial SLI is at $z=z_0$, we partition the $z=z_0$ plane into $n_g$ regions. For phase field $\alpha$, if point $(x,y)$ is in region $\Omega_{\alpha}$ we use a $\tanh$ function to create a smooth transition from solid (1) to liquid (-1) in the $z$-direction; if point $(x,y)$ is outside of $\Omega_{\alpha}$, we set $\phi_{\alpha}$ to -1. $\ksiz$ parameterizes the initial realization of $\Omega_{\alpha}$. Here we use the Voronoi diagram \cite{CHADWICK2021116862} which creates the Voronoi tessellations from a set of seed points. We let $\ksiz = \{\boldsymbol{x}^0_j\}^{n_j}_{j=1}\cup\{\boldsymbol{\theta}_g\}^{n_g}_{g=1}$, where $\boldsymbol{x}^0_j$ are the vertices of the Voronoi diagram and $\boldsymbol{\theta}_g$ are the grain orientation for each phase field. With a crystal orientation angle $\boldsymbol{\theta}_g$, the anisotropy functions \Cref{eqn:anis0} and \Cref{eqn:anis1} are modified by replacing all spatial derivatives by the derivatives with respect to the rotated coordinate system ($x^{\prime}$, $y^{\prime}$, $z^{\prime}$) with angle $\boldsymbol{\theta}_g$ \cite{Tourret2015a}. For each phase field simulation, $\ksiz$ is randomly generated. The orientation of each grain is sampled from the unit sphere. If $\boldsymbol{d}$ is the vector representing the grain's orientation, we first sample  $\boldsymbol{d} \sim \mathcal{N}(\boldsymbol{0},\boldsymbol{I})$ and set  $\boldsymbol{d}=\boldsymbol{d}/|\boldsymbol{d}|_2$. Samplings of $\boldsymbol{x}^0_j$ are discussed in \Cref{sec:training}.




\Cref{fig:1phase_field} shows an example of a simulation with domain size $(120 \mu m, 120 \mu m, 50\mu m)$. The interface width $W_0=0.1\mu m$ and the mesh size $dx = 0.08\mu m$ \cite{CHADWICK2021116862}. Therefore the grid size is $1500 \times 1500 \times 625$.  The initial conditions for grain microstructure in the substrate are generated by sampling $\boldsymbol{x}^0_j$ from a uniform distribution and using a Voronoi diagram with periodic boundary conditions. In \Cref{fig:1phase_field}a, the substrate has 900 grains.  In this simulation $G_z$ and $R_z$ are $10K/\mu m$ and $2m/s$, respectively. The time step size is 2.42 nanoseconds. \Cref{fig:1phase_field}b is the resulting phase field microstructure after 10K time steps, which required approximately 48 minutes on one Nvidia A100 GPU. 

Given the grain microstructure, we compute several \emph{quantities of interest (QoIs)}: number (or percentage) of eliminated grains; grain size distribution; and volume-averaged misorientation. The number of eliminated grains is relevant in epitaxial growth \cite{takaki2018competitive}. We compute $n_G=n_G(z)$, the accumulated grain eliminations from the initial interface location to a height $z$ (see \Cref{fig:1phase_field}f). Grain size distribution is a common descriptor of grain shape statistics and it affects mechanical properties such as strength and ductility. Grain size $d$ of grain $g$ is defined by its volume-equivalent diameter $d_g=(6\mathcal{V}_g/\pi)^{1/3}$, where $\mathcal{V}_g$ is the grain volume. \Cref{fig:1phase_field}d shows the probability distributions of the grain size for \Cref{fig:1phase_field}a and \Cref{fig:1phase_field}b. Volume-weighted misorientation $\Delta \boldsymbol{\theta}$ quantifies the alignment of the polycrystalline orientation with the prescribed temperature gradient direction $z$. $\Delta \boldsymbol{\theta} = \sum_g \mathcal{V}_g \boldsymbol{\theta}_{z,g}/\sum_g \mathcal{V}_g$ and $\boldsymbol{\theta}_{z,g}$ is the angle between the crystal orientation of grain $g$ and the $z$-axis. As shown in \Cref{fig:1phase_field}f, $\Delta \boldsymbol{\theta}$ decreases as the more aligned grains out-compete the misaligned grains. Due to random initial condition $\ksiz$, we need to average across several simulations to obtain expectations of the QoIs.

\begin{figure}[h!]
 \centering
     \includegraphics[width=1\textwidth]{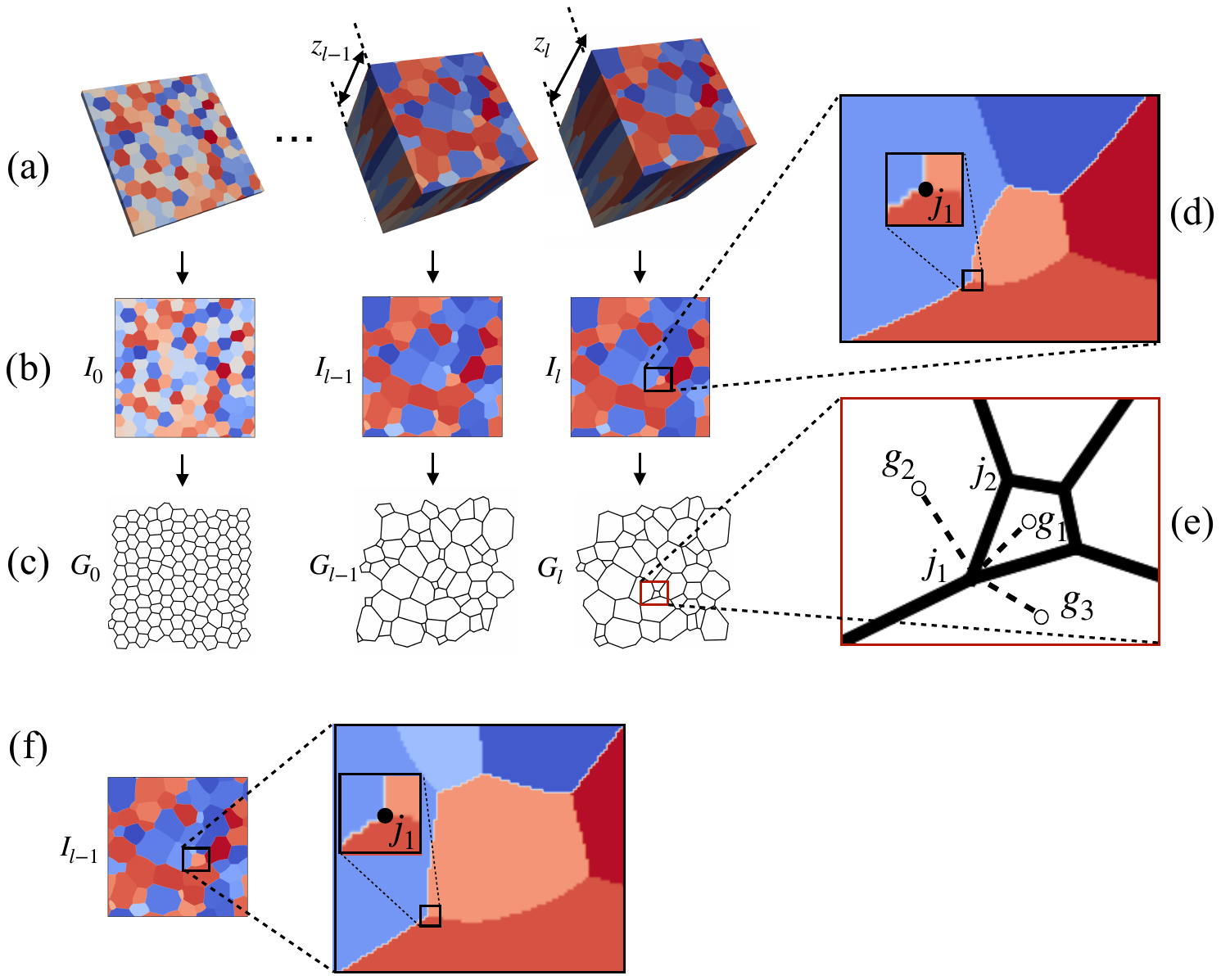}
     \caption{\textbf{Graph data extraction from phase field simulations.} (a) Snapshots of a phase field simulation. Each snapshot is taken when the solid-liquid interface reaches height $z_l=l\Delta z, l=0,1,2,...$ (b) $I_l$ is a 2D image of grain index field at height $z=z_l$. (c) Graph $G_l$ extracted from the image $I_l$. (d) Zoomed-in inset of a region in the image $I_l$: $j_1$ is a junction pixel with three neighboring grains, $g_1,g_2,g_3$. (e) The same region in the graph $G_l$: $G_l$ has two kinds of vertices and two kinds of edges; $j_1$ is a junction vertex corresponding to the junction pixel in (d); $g_1$ is a grain vertex; $(j_1, j_2)$ is a junction-junction edge; $(j_1, g_1)$ is a junction-grain edge. (f) A zoomed-in region in image $I_{l-1}$: $j_1$ is a junction with the same grain neighbors as $j_1$ in (d). We compare the grain neighbors to identify the same junctions in different graphs.}
     \label{fig:2graph}
\end{figure}

\subsection{GrainGNN}\label{sec:grainGNN}
The input to GrainGNN is the initial state $\ksiz$ of the substrate  (see \Cref{fig:1phase_field}a), process parameters $\piz=\{G_z, R_z\}$, and a domain size $L=(L_x, L_y, L_z)$. 
The output of GrainGNN is the grain orientation field $\boldsymbol{\theta}(x,y,z)$ in a domain with size $L$. Notice that here $L_z$ is the height of the final polycrystal (as shown in \Cref{fig:1phase_field}b); for phase field simulations we let $L^{\prime}_z>L_z$ to avoid the effect of the top boundary. In sections \Cref{sec:2graph}-\Cref{sec:5graph_update} we explain the basic ingredients of GrainGNN. 
\begin{itemize}

\item {\bf Image-to-graph step}. We start with a spatial field compression method to reduce the large dimensionality of the phase field data. We define a planar graph $G$ and hand-crafted features $F$ to represent a layer of grain microstructure (see \Cref{fig:2graph}). To define these features from $\{\phi_{\alpha}\}_{\alpha=1}^{n_g}$ we first combine the fields to a single image $\boldsymbol{\theta}(x,y,z)$.
  Then we propose an image-to-graph algorithm $\phi_{\alpha}(x,y,z) \rightarrow (G_l, F_l), \ l=0,1,2,...$ to extract vertices and edges of grains at cross-sections defined at $z_l$. The features contain geometric information about the size of the grains and their boundary.  The definitions of the image, graph, and features are discussed in \Cref{sec:2graph}.

\item {\bf Graph evolution step}. We now train a network for the map $(G_{l-1},F_{l-1})\rightarrow(G_l,F_l)$. The challenge is that graphs $G_{l-1}$ and $G_{l}$ can be topologically different due to grain and/or edge elimination events. To address this we need to predict both vertex feature evolution  $\Delta F$ as well as a set of \emph{grain neighbor switching events} $\mathcal{S}_E$ and a set of \emph{grain elimination events} $\mathcal{S}_G$.   We discuss the graph evolution in detail in \Cref{sec:3graph_dynamics}. The prediction $(G_l,F_l)$ is done by first using GrainGNN to predict $\Delta F$ and \emph{probabilities} of events and then using an algorithm to reconstruct $G_l$ and its features.
  \begin{itemize}
  \item {\bf Graph evolution, LSTM substep.} We design an LSTM  architecture to learn the predictions $(\Delta F, \mathcal{S}_E, \mathcal{S}_G)= \text{GrainGNN} (G_{l-1}, F_{l-1})$. GrainGNN consists of two networks $\mathcal{R}$ and $\mathcal{C}$.  $\mathcal{R}$ is a regressor that outputs $\Delta F$ and $\mathcal{S}_G$; $\mathcal{C}$ is a classifier that predicts $\mathcal{S}_E$.  We present the summary of the network architecture \Cref{fig:4graingnn} and the details in \Cref{sec:4arch}. 

  \item {\bf Graph evolution, graph reconstruction substep.} Given the LSTM predictions we introduce a graph update algorithm that completes the graph prediction $(G_{l-1}, \mathcal{S}_E, \mathcal{S}_G) \rightarrow G_l$, where we create orderings for $\mathcal{S}_E$ and $\mathcal{S}_G$ to reconstruct the graph $G_l$ (see \Cref{fig:4update}). The features are updated $(F_{l-1}, \Delta F, \mathcal{S}_E, \mathcal{S}_G) \rightarrow F_l$. We give the details of the graph reconstruction algorithm in \Cref{sec:5graph_update}.
    \end{itemize}

\item {\bf Graph-to-image, microstructure reconstruction.} Using GrainGNN we  can compute a graph ``trajectory'' $\mathcal{G}=\{ G_0, G_1, G_2...,\}, \mathcal{F}=\{ F_0, F_1, F_2...,\}$, given an initial condition: $(G_0, F_0)\rightarrow (\mathcal{G}, \mathcal{F})$. Once the graph trajectory is computed, we introduce a graph-to-image algorithm $(\mathcal{G}, \mathcal{F}) \rightarrow \boldsymbol{\theta}(x, y, z)$ to reconstruct the grain orientation field; the algorithm is described in \Cref{sec:6graphTOimage}.
\end{itemize}

\subsection{Graph representation of microstructure}
\label{sec:2graph}

Now, let us describe the compressed representation of the microstructure using hand-crafted features. Consider a solution of~\Cref{eq:micro_jsi}. From this simulation we sample  $n_l$ layers of field data. 
We define a ``layer'' at height  $z_l$ as the microstructure at the time $t_l$ when the solid-liquid interface (SLI) reaches  $z_l$ (see \Cref{fig:2graph}a). More precisely, $t_l$ is defined as the time at which the lowest point (order in $z$ coordinate) at the SLI reaches $z_l$.

We define $I_l(x,y)$ to be an \emph{image}, a 2D scalar representation of the microstructure on the $x-y$ plane at $z_l$. 
Each layer $I_l$ is a 2D image on the $x$-$y$ plane (\Cref{fig:2graph}b): 

\begin{equation}
I_l(x,y) = \underset{\alpha}{\arg \max }\ \phi_{\alpha}(x,y,z_l,t_l).
\label{eqn:grain_index}
\end{equation}
We solve \Cref{eq:micro_jsi} to generate training data and verify the predictions of the surrogate. Although a simulation can have thousands of time steps, we subsample to $n_l$ planes using equispaced sampling for $z_l$ with height increment $\Delta z$, so $z_l=l\Delta z$. Typically we use $n_l=20$. We discuss the choice of $\Delta z$ and $n_l$ in \Cref{sec:3graph_dynamics}. 

\paragraph{Definition of edges and vertices.}
Once $I_l$ have been identified, we extract  graph $(G_l,F_l$ (\Cref{fig:2graph}c) as follows.
$G(V, E)$ is an undirected  graph where $V$ and $E$ denote the sets of vertices and edges respectively.  In GrainGNN, $G$ has two types of vertices $V=V_{g}\cup V_{j}$. $V_{g} = \{ {g}_1, g_2,..., {g_{n_g}} \}$ represents grains. $V_{\boldsymbol{j}} = \{ {j}_1, {j}_2,..., {j_{n_j}} \}$ are junction vertices, where $n_j$ is the number of junctions. A junction vertex refers to a point in image $I_l$ that has three grain neighbors (e.g., $j_1$ in \Cref{fig:2graph}d). Notice that the junction vertices are just the vertices of the Voronoi diagram. $G$ has two types of undirected edges $E=E_{jj} \cup E_{jg}$. $E_{jj}$ are the edges between junction vertices and can be thought of as a representation of actual inter-grain boundaries on the $x$-$y$ plane. $E_{jg}$ are the junction-grain edges (dashed lines in \Cref{fig:2graph}e).

As we will see later we need to define a maximum degree (number of edges) for a junction vertex.  The maximum number of edges for a junction vertex depends on the physical setting of the problem. In rapid solidification of metals, the grain boundary anisotropy is relatively weak compared to kinetic anisotropy~\cite{bragard2002linking}, and a triple junction is more stable than a quadruple junction \cite{torres2015numerical}. Thus we only consider triple junctions in this work, which means a junction vertex can only connect to the other three junction vertices and three grain vertices. In this case, $n_j=2n_g$, $|E_{jj}|=3n_g$, and $|E_{jg}|=6n_g$. Our handling of the graph topological changes discussed in \Cref{sec:3graph_dynamics} is based on this simplification of the graph structure. 

Let's now discuss how to identify vertices and edges from $I_l$. First, we identify the junction locations $(x_j$, $y_j)$. For each pixel $(x, y)$ of $I_l$, we inspect its eight neighbors each associated with a grain index. If there are three distinctive grain indices in eight neighbors, we consider pixel $(x, y)$ to be a junction pixel, for example, $j_1$ in \Cref{fig:2graph}d. Every junction $j$ is associated with a unique triplet of grain indices $N_j=\{g_1,g_2,g_3\}$. Commonly multiple adjacent pixels will have the same triplet and we only keep one of them\footnote{If two or more pixels have the same triplet $\{g_1,g_2,g_3\}$. We count the occurrences of each grain index in the eight neighbors of each pixel and use the one pixel whose index occurrences are more even. For example, for pixel 1 the occurrences are $\{ 3, 3, 2\}$ and for pixel 2 the occurrences are $\{ 4, 3, 1\}$ and we choose pixel 1.}. 
Then using the triplet indices $N_j$ we add three edges to the $E_{jg}$ set: e.g., for $j_1$,  the $E_{jg}$ edges are $({j_1}, g_1)$, $({j_1}, g_2)$, $({j_1}, g_3)$. We create $E_{jj}$ edges by  $E_{jj}=\{(j_1,j_2):N_{j_1} \cap N_{j_2} =2, \forall j_1, j_2 \in G\}$. For example in \Cref{fig:2graph}e, edge $(j_1, j_2)$ exists because $j_1$ and $j_2$ share the same neighboring grains $g_1$ and $g_2$. 

\paragraph{Definition of features.}
We clarify that these are input features to the LSTM and serve as a reduced representation of the grain structure. They are not the network hidden features, which are of course determined during training.  We define vertex and edge features as follows:
\begin{itemize}
  \item {\bf Junction vertices:} The feature vector of a junction vertex is defined as:
\begin{equation}
\boldsymbol{f}_j = \frac{1}{\boldsymbol{f}^0_j}{\begin{bmatrix} x_j & y_j & z_l & G_z & R_z & \Delta x_j & \Delta y_j & \Delta z \end{bmatrix}}, 
\label{eqn:junction_feature}
\end{equation}
where $x_j$, $y_j$, and $z_l$ are the 3D coordinates of a junction vertex; $\Delta x_j$ and $\Delta y_j$ are in-plane displacements of junctions with respect to their locations in the previous layer $G_{l-1}$, i.e., $\Delta x_j=x_{j,l}-x_{j,l-1}, \Delta y_j=y_{j,l}-y_{j,l-1}$. $\Delta z=z_l-z_{l-1}$ is the distance between two sampled images. For $l=0$, we set $\Delta x_j = \Delta y_j = \Delta z = 0$.  $G_z$ and $R_z$ here are constants; they're repeated at each vertex and we allow them to vary for each junction location.

  \item {\bf Grain vertices:} For a grain vertex, we define $\boldsymbol{f_g}$ as: 
\begin{equation}
\boldsymbol{f}_g = \frac{1}{\boldsymbol{f}^0_g}\left[ \begin{array}{@{}*{11}{c}@{}}
     x_g & y_g & z_l & s_g & v_g& \cos\boldsymbol{\theta}_x & \sin\boldsymbol{\theta}_x & \cos\boldsymbol{\theta}_z & \sin\boldsymbol{\theta}_z & \Delta s_g & \Delta z \\
\end{array} \right],
\label{eqn:grain_feature}
\end{equation}
where $x_g$, $y_g$, and $z_l$ are the coordinates of the grain vertices; $s_g$ is the grain cross-sectional area; and $v_g$ is the grain excess volume above the interface height $z_l$ \cite{qin2023grainnn}. For $x_g$, we average the coordinates of its junction neighbors $x_g = \sum_{k \in N_g} x_{j,k}/|N_g|$.  $N_g$ is the set of junctions connected to grain $g$.
$\boldsymbol{\theta}_z$ is the angle between the $z$-axis and the preferred growth direction <100>; $\boldsymbol{\theta}_x$ is the angle between the $x$-axis and the projection of <100> direction on the $x$-$y$ plane.
  $\Delta s_g = s_{g,l} - s_{g,l-1}$ is the change of cross-sectional area; $\Delta s_g =0$ at $l=0$. In \Cref{eqn:grain_feature} and \Cref{eqn:junction_feature}, $\boldsymbol{f}^0_j$ and $\boldsymbol{f}^0_g$ are constant vectors to normalize features to [0, 1]. $\boldsymbol{f}^0_j=\begin{bmatrix} L_x, L_y, L_z, G_{\mathrm{max}}, R_{\mathrm{max}}, L_x, L_y, L_z\end{bmatrix}$ and $\boldsymbol{f}^0_g=\begin{bmatrix} L_x, L_y, L_z, L_xL_y, L_xL_yL_z, 1,1,1,1, L_xL_y, L_z\end{bmatrix}$. $G_{\mathrm{max}}$ and $R_{\mathrm{max}}$ are maximum values of $G_z$ and $R_z$ in the training data.

\item {\bf Edge features:} For each $E_{jj}$ and $E_{jg}$ edge, we include its length as the edge feature. Notice the length of an edge is calculated with periodic boundary conditions. We neglect the curvature of $E_{jj}$ edge for simplicity (but notice that the grain boundary curvature in the $z$-direction is still captured). 
\end{itemize}
In summary, $F$ contains three feature matrices $F_j=\begin{bmatrix} \boldsymbol{f}_{j,1} & \boldsymbol{f}_{j,2} & ... & \boldsymbol{f}_{j, n_j} \end{bmatrix}$, $F_g=\begin{bmatrix} \boldsymbol{f}_{g,1} & \boldsymbol{f}_{g,2} & ... & \boldsymbol{f}_{g, n_g}  \end{bmatrix}$, and edge features $F_E=\{|\boldsymbol{x}_i - \boldsymbol{x}_j|, \forall (i,j) \in E \}$.
The values of  $\Delta x_j$, $\Delta y_j$, $\Delta s_g$ are obtained by subtracting $F_l$ and $F_{l-1}$, which is discussed in the next section.

\begin{figure}[h!]
 \centering
     \includegraphics[width=0.85\textwidth]{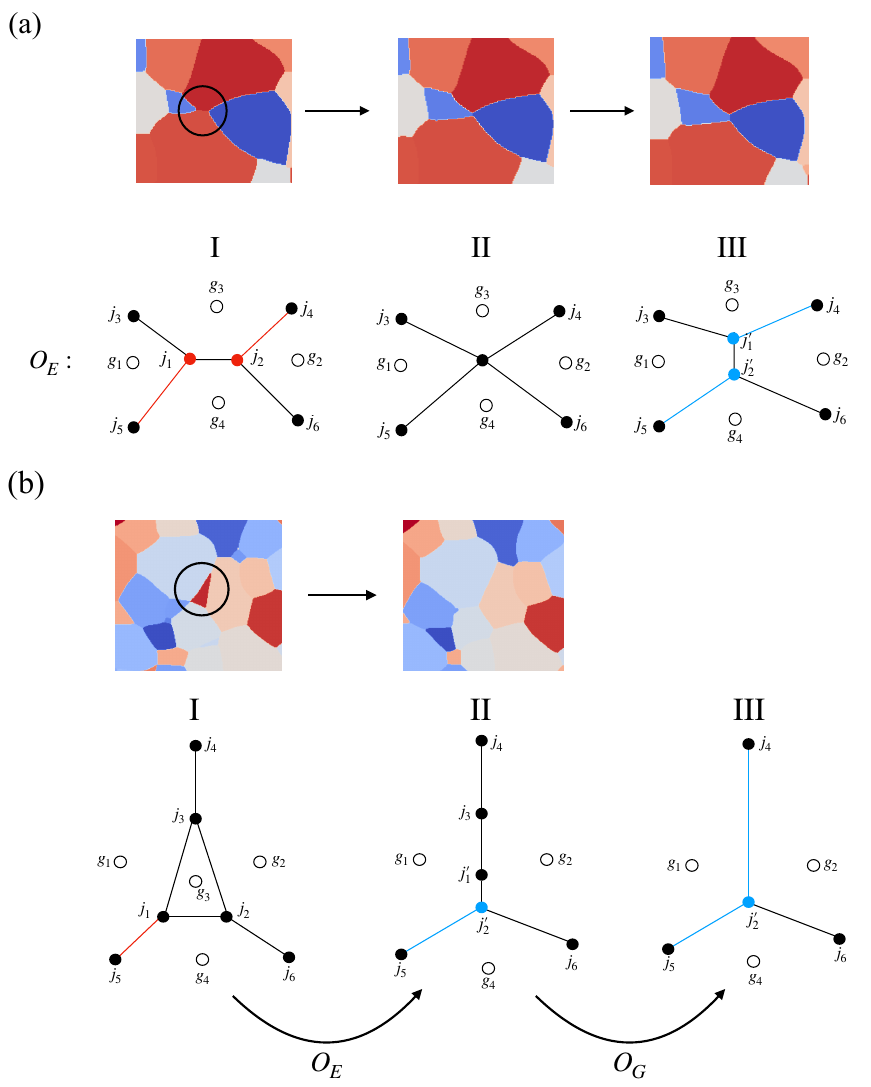}
     \caption{\textbf{Topological events with the update of graph connectivity.} (a) A neighbor-switching event requires changes of the graph topology: an $O_E$ operation. Grains $g_3$ and $g_4$ initially are neighbors. After  applying $O_E$, $g_1$ and $g_2$ are neighbors. From \rom{1} to \rom{2}, $j_1, j_2$ merge to a quadruple junction; from \rom{2} to \rom{3}, the quadruple junction  splits into two triple junctions. This event replaces edges $(j_1, j_5)$, $(j_2, j_4)$ (in red color) with the edges $(j^{\prime}_2, j_5)$, $(j^{\prime}_1, j_4)$ (in blue color). The total number of edges remains the same. (b) A grain elimination event. From \rom{1} to \rom{2}, a neighbor-switching event occurs on $(j_1, j_2)$; $g_3$ becomes a 2-side grain. \rom{2} to \rom{3} is an $O_G$ operation. It removes $g_3$, $j^{\prime}_1$, $j_3$, a couple of edges, and replace them with $(j^{\prime}_2, j_4)$. }
     \label{fig:3events}
\end{figure}

\subsection{Graph evolution}
\label{sec:3graph_dynamics}

We now present how to model the graph-to-graph mapping $(G_{l-1}, F_{l-1}) \rightarrow (G_l, F_l)$. The dynamics observed in epitaxial grain growth can be decomposed into three basic types \cite{zollner2017topological, torres2015numerical}: (i) junction vertex movement, (ii) grain neighbor switching, and (iii) grain elimination.  We remark that (i) does not change the graph, only the value of its features; (ii) changes features and it removes and adds edges. (iii) does both (i) and (ii) but also removes vertices, both junction and grain vertices. We define two basic operations that compose the topological changes from $G_{l-1}$ to $G_l$. One is a neighbor-switching operation $O_E$; another is a vertex-removal operation $O_G$. By operation we mean a sequence of deterministic actions that add/remove vertices or edges. One neighbor switching event causes one $O_E$; one grain elimination event causes one $O_G$ and multiple $O_E$. Next we discuss these operations in detail.

{\bf Grain neighbor switching events:} Grain neighbor switching happens when a grain edge becomes smaller and eventually disappears. A disappearing edge creates a four-edge junction (\Cref{fig:3events}a). Since in our PDE model four-edge junctions are unstable, a new edge starts growing and the four-junction becomes two triple junctions. This new edge causes the neighbor to switch.  Physically, a neighbor-switching event involves four grains, where two grains lose one face and the other two grains gain one face. From \rom{1} to \rom{2}, grain $g_1$ and $g_2$ push the vertices $j_1$ and $j_2$ to move toward each other. At some intermediate time, $j_1$ and $j_2$ merge to form a quadruple junction.
Then from \rom{2} to \rom{3}, the quadruple junction separates into two new junctions $j^{\prime}_1$ and $j^{\prime}_2$, which replace $j_1$ and $j_2$. This event changes several edges, grain $g_1$ and $g_2$ grain an additional junction-grain edge; $g_3$ and $g_4$ lose one junction-grain edge; $(j^{\prime}_2, j_5)$ and $(j^{\prime}_1, j_4)$ replace $(j_1, j_5)$ and $(j_2, j_4)$; the total number of edges remains the same. The grain neighbor-switching event is characterized by the removal of $(j_1, j_2)$. We will refer to a grain neighbor switching event as an ``edge'' event in the following discussion.

{\bf Grain elimination events:}
\Cref{fig:3events}b showcases a grain elimination event on $g_3$ which has three grain neighbors. In this example, face $(j_1, j_2)$ shrinks faster than two other faces and triggers a neighbor switch with $j^{\prime}_1$, $j^{\prime}_2$ the new junctions. \rom{2} to \rom{3} is an $O_G$ operation. In this operation, we remove vertices $g_3$, $j^{\prime}_1$, $j_3$, and edges connected to $j^{\prime}_1$ and $j_3$. Finally, we add a new edge $(j^{\prime}_2, j_4)$. Note that one grain about to be eliminated can have a different number of faces, typically three to seven. Elimination of an $|N_g|$-side grain needs $|N_g|-2$ $O_E$ operations and one $O_G$ operation. One $O_G$ operation removes one $V_g$ vertex, two $V_j$ vertices, three $E_{jj}$ edges, and six $E_{jg}$ edges.

{\bf Matching $G_{l-1}$ and $G_l$ and handling elimination events:}
Our graph update requires computing  $\Delta F, \mathcal{S}_E, \mathcal{S}_G$. We define
\begin{equation}\label{e:deltaF}
\Delta F = \{\Delta x_j, \Delta y_j, \Delta s_g, v_{g,l}, \ \forall j, g \in G_{l-1} \}.
\end{equation}
To create $\Delta F$ for training data, we first need to match the vertices between $G_{l-1}$ and $G_l$.

If there were no topological changes, the two graphs can be easily matched as follows. $V_g$ are identical for the two graphs. Junction vertices are matched by their grain index triplets $N_j$. For example in \Cref{fig:2graph}d and \Cref{fig:2graph}f, the magnified junctions have the same grain neighbors so they are the same junction vertex. $\Delta x_j, \Delta y_j, \Delta s_g$ are then obtained by subtracting features of their matched vertices.

Grain elimination events $\mathcal{S}_G$ can also be handled relatively easily. They cause grain vertices in $G_{l-1}$ to be missing from $G_{l}$. These vertices are defined by the set  $\mathcal{S}_G= \{g: g \in G_{l-1} \wedge g\notin G_{l}\}$.

\begin{figure}[h!]
 \centering
     \includegraphics[width=1\textwidth]{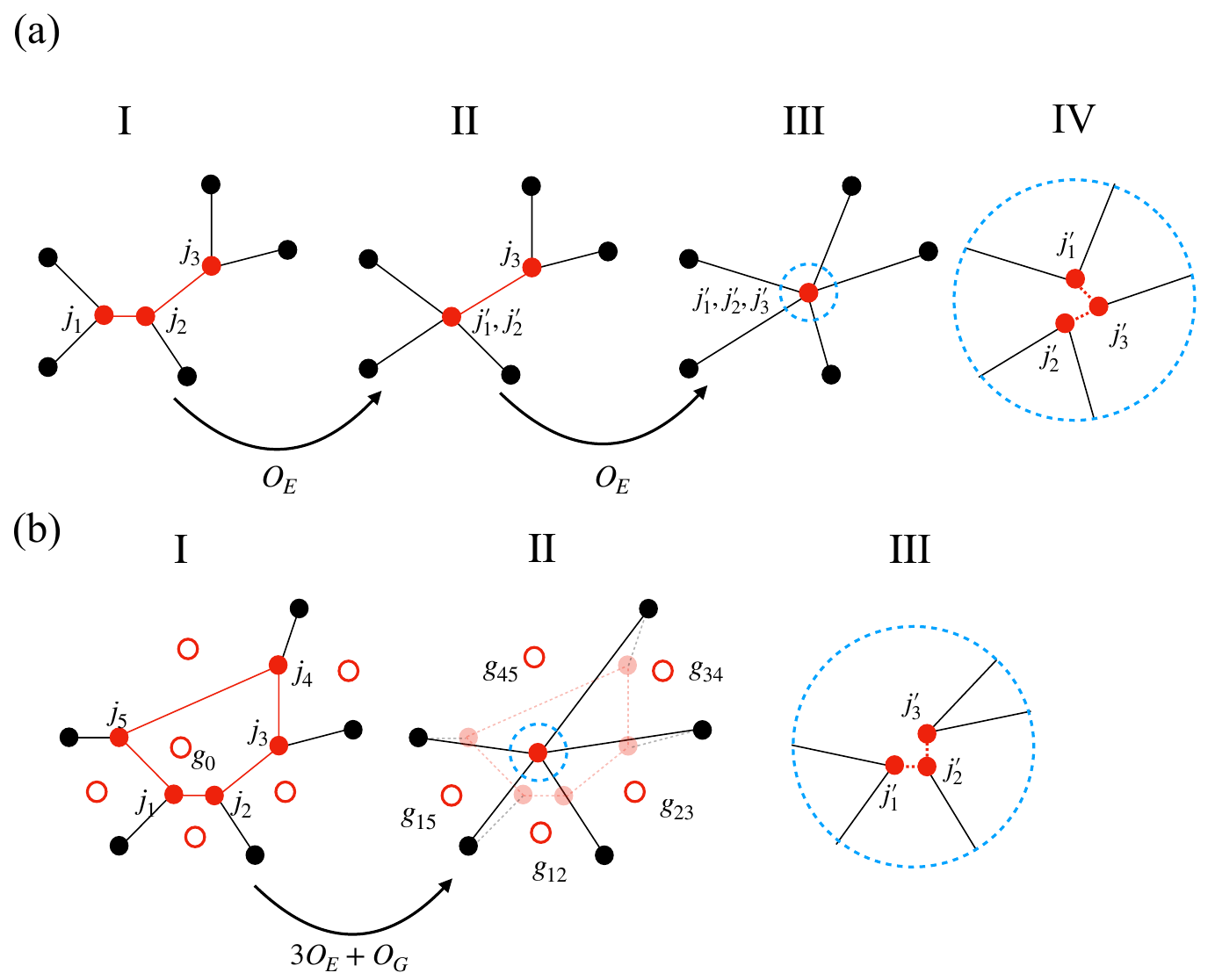}
     \caption{\textbf{GrainGNN graph update orderings given events lists $\mathcal{S}_E$ and $\mathcal{S}_G$.} (a) Order of edge events. $(j_1, j_2)$ and $(j_2, j_3)$ are edges with elimination probability $P_{12}>P_{23}>\epsilon_E$.  GrainGNN applies $O_E$ on $(j_1, j_2)$ before $(j_2, j_3)$. For each $O_E$, the edge connectivity is updated as shown in \Cref{fig:3events}a. We assume new junction points $j^{\prime}_1,j^{\prime}_2$ are both located at the midpoint of edge $(j_1, j_2)$. After event $(j_2, j_3)$, $j_1, j_2, j_3$ are at the same location but the connectivity is as shown in IV.(b) Elimination of grain $g_0$. The list of $O_E$ operations is sorted by the area increment of neighboring grains. Here $\Delta s_{g_{15}}<\Delta s_{g_{12}}<\Delta s_{g_{23}}<\Delta s_{g_{45}}<\Delta s_{g_{34}}$. Thus GrainGNN applies $O_E$ in the order $(j_1,j_5), (j_1,j_2), (j_2,j_3)$.}
     \label{fig:4update}
\end{figure}

Neighbor-switching events are harder to handle: they result in junctions whose triplets $N_j$ are different in $G_l$ and $G_{l-1}$. Let $\mathcal{U}_{l-1}$ be the junction vertices in $G_{l-1}$ but not matched in $G_{l}$, $\mathcal{U}_{l-1}= \{j: j\in G_{l-1} \wedge j\notin G_{l}\}$. We enumerate every two junctions in $\mathcal{U}_{l-1}$ and check (i) if they have an $E_{jj}$ edge and (ii) if they become two new vertices of $G_l$ through a neighbor switching event. For the second check, we use their $N_j$ triplets. We use \Cref{fig:3events}a as an example. $N_{j_1} = \{g_1, g_3, g_4\}$ and $N_{j_2} = \{g_2, g_3, g_4\}$. If an edge event happened on $(j_1, j_2)$, two new triplets $N_{j^{\prime}_1} = \{g_1, g_2, g_3\}$ and $N_{j^{\prime}_2}=\{g_1, g_2, g_4\}$ must have formed at an intermediate time. We check that if $N_{j^{\prime}_1}$ and $N_{j^{\prime}_2}$ exist in $G_l$, criteria (ii) is satisfied. Thus neighbor-switching events $\mathcal{S}_E$ are junction pairs in $\mathcal{U}_{l-1}$ that satisfy (i) and (ii). We remark this method cannot match all the junction vertices. When a junction has been involved in two or more edge or grain elimination events  (e.g., \Cref{fig:4update}a), we fail to detect the events. In this case, we mask out (in training) the unmatched junctions and $E_{jj}$ edges. These are typically less than 1\% of the total vertices and edges.

{\bf Choosing $\Delta z$:} Finally, we discuss how to choose $\Delta z$, the distance between $I_{l-1}$ and $I_{l}$. If $\Delta z$ is too small, GrainGNN will need too many steps during inference, and make the surrogate too expensive.  If $\Delta z$ is too large, $I_{l-1}$ and $I_{l}$ will be exceedingly different as they will involve a large number of topological events. With these observations in mind, we choose the number of layers $n_l$ such that:
\begin{equation}
\begin{aligned}
     & \frac{n_G(z=L_z)}{n^0_g n_l} \approx 3\%, \\
     & \Delta z = L_z/(n_l-1),   
\end{aligned}
\label{eqn:num_layer}
\end{equation}
where $n^0_g$ is the number of grain vertices of $G_0$. We allow about 3\% of grains to be eliminated per graph update. Notice that $n_G<n^0_g$ so $n_l \sim O(1)$. For training data, we count $n_G$ and calculate $n_l$ and $\Delta z$ with \Cref{eqn:num_layer}. For example, if we run a 100-grain phase field simulation to have 40 grains at the end, $n_l = (100-40)/100/3\%=20$. For GrainGNN inference, we don't know $n_G$ beforehand. We first create map $\Delta z(\piz)$ using the training $\piz$ grid. Then for a testing $\piz$, we use the nearest neighbor interpolation for $\Delta z$.

\begin{figure}[h!]
 \centering
     \includegraphics[width=1\textwidth]{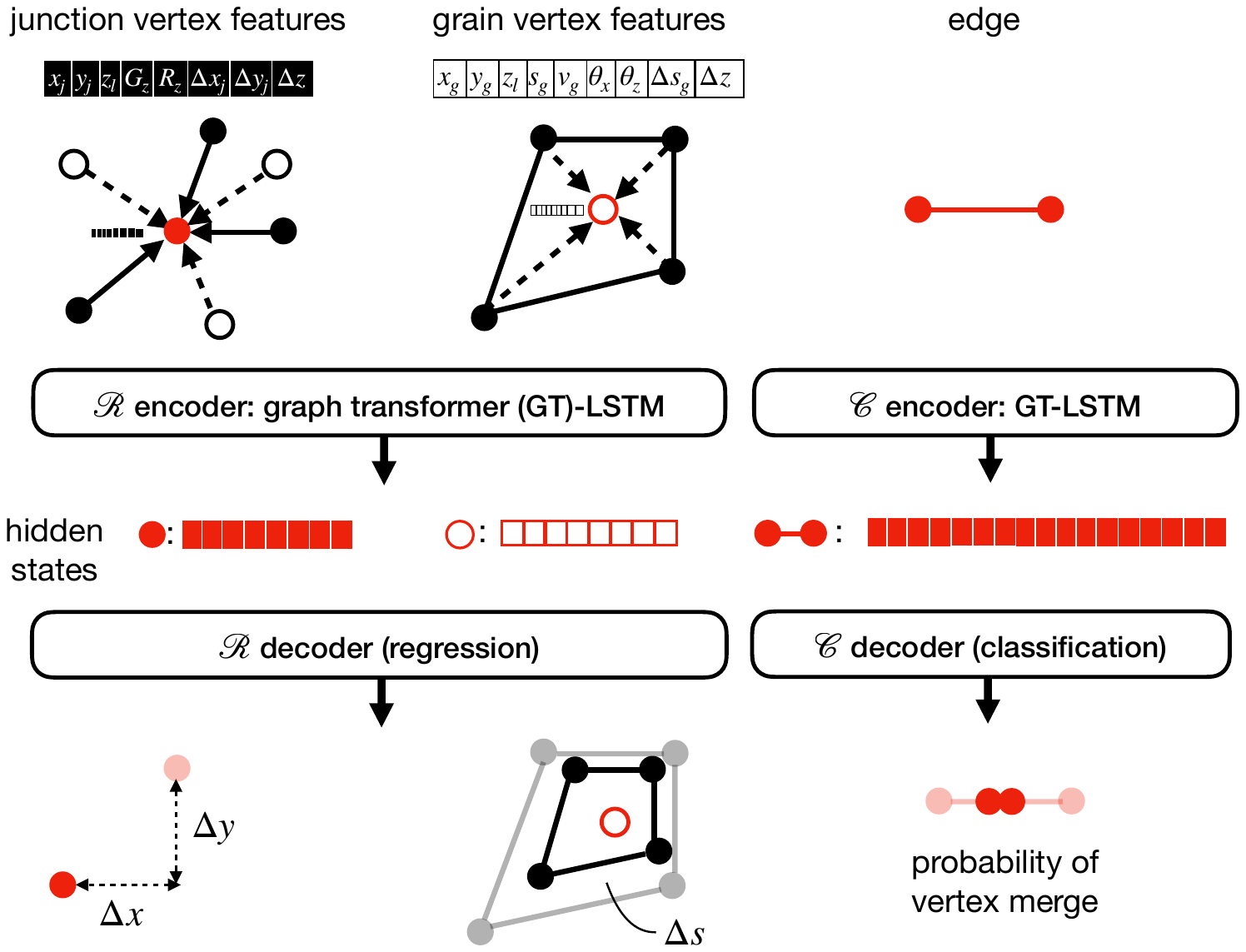}
     \caption{\textbf{The LSTM architecture.} We use a graph transformer LSTM with  a regressor $\mathcal{R}$ and a classifier $\mathcal{C}$---both having an  encoder-decoder structure. The classifier and regressor are trained separately. The junction nodes (solid circles) and grain nodes (hollow circles) have vertex features defined in \Cref{eqn:junction_feature} and \Cref{eqn:grain_feature}. For each target vertex (red circles), graph transformer operators in \Cref{eqn:lstm} aggregate the features of the vertex itself with the neighboring vertices into a hidden vector. The decoder $\mathcal{R}$ then uses \Cref{eqn:r_decoder} to transform the hidden vector to the target outputs, which are displacements of junctions, area change, and excess volume of grains. For each junction-junction edge, the model $\mathcal{C}$ concatenates the hidden vectors of the two connecting vertices and predicts the probability of the edge event (see \Cref{eqn:c_decoder}). }
     \label{fig:4graingnn}
\end{figure}

\subsection{The LSTM architecture}
\label{sec:4arch}

In this section, we discuss the neural network at the heart of the GrainGNN surrogate. It comprises a regressor $\mathcal{R}$ and a classifier $\mathcal{C}$. Both have an encoder-decoder structure. Graph features of $F_{l-1}$ are encoded to intermediate hidden states $H_{l}$ through the encoder and then $H_{l}$ are decoded to the target outputs (see \Cref{fig:4graingnn}). We use a graph transformer LSTM as the encoder for both $\mathcal{R}$ and $\mathcal{C}$. The encoders for $\mathcal{R}$ and $\mathcal{C}$ have the same architecture but they have different weights and they're trained separately.

{\bf LSTM encoder architecture.}
Each encoder has several identical LSTM \cite{hochreiter1997long} layers stacked on top of each layer. The  LSTM layer hidden states are denoted by  $H \in \mathbb{R}^{ D_h \times ( n_j + n_g )}$; its  cell states by  $C \in \mathbb{R}^{ D_h \times ( n_j + n_g )}$. Here $D_h$, the hidden dimension, is a network architecture hyperparameter. Hidden and cell states are intermediate outputs of LSTM to store short- and long-term prior information. The inputs to each LSTM layer are the feature matrices $F_{l-1}$ and the graph adjacency matrix $A_{l-1}$ ($A_{ij}=\mathbb{1}_{(i,j)\in E}$). The outputs are the updated hidden and cell states $H_l$, $C_l$. Let $U_{l-1}=[F_{l-1}, H_{l-1}]$ be the input matrix, and one LSTM layer is defined as follows: 
\begin{equation}
\label{eqn:lstm}
\begin{aligned} 
i_{l} &=\sigma\left(\mathcal{T}_{i}\left(U_{l-1}, A_{l-1}\right)+b_{i}\right), \\ 
f_{l} &=\sigma\left(\mathcal{T}_f\left(U_{l-1}, A_{l-1}\right)+b_{f}\right), \\ 
C_{l} &=f_{l}  C_{l-1}+i_{l}  \tanh \left(\mathcal{T}_c\left(U_{l-1}, A_{l-1}\right)+b_{c}\right), \\
o_{l} &=\sigma\left(\mathcal{T}_o\left(U_{l-1}, A_{l-1}\right)+b_{o}\right), \\ 
H_{l} &=o_{l}  \tanh \left(C_{l}\right).
\end{aligned}
\end{equation}
where $\mathcal{T}_{i}$, $\mathcal{T}_f$, $\mathcal{T}_c$, $\mathcal{T}_o$ are graph transformer operators \cite{shi2020masked} with trainable weights; $\sigma$ is the sigmoid function; $b_i$, $b_f$, $b_c$, $b_o$ are biases. For the first layer, $H_{l-1}$ and $C_{l-1}$ are initialized as zeros; for other layers, $H_{l-1}$ and $C_{l-1}$ are initialized with $H_l$ and $C_l$ of the previous layer. 

The graph transformer operator $\mathcal{T}$, is a  message-passing neural network \cite{duvenaud2015convolutional}. $\mathcal{T}$ operates on every node in the graph with the node's neighboring nodes. In \Cref{fig:4graingnn} (top row), the red circles indicate a node on which we evaluate $\mathcal{T}$ and the black circles its neighbors. Every neighbor passes its vertex and edge features 
to the target node. $\mathcal{T}$ aggregates the passed information and the features of the target node into a single vector. Let $\boldsymbol{u}_{i}=[\boldsymbol{f}_{i}\ \boldsymbol{h}_{i}]$ be the input vector of node $i$; $\boldsymbol{h}_i \in \mathbb{R}^{D_h}$ is node $i$'s hidden vector. 
$\mathcal{T}(U, A)_i \in \mathbb{R}^{D_h}$, the output vector for node $i$, is given by \cite{shi2020masked}:
\begin{subequations}
\label{eqn:GT}
\begin{align}
& \mathcal{T}(U, A)_i=W_{1} \boldsymbol{u}_{i}+\sum_{k \in N_i} \beta_{i, k}\left({W}_{2} \boldsymbol{u}_{k,i}+{W}_{3} \boldsymbol{f}_{ik}\right), \\
& \beta_{i,k}=\operatorname{softmax}\left(\frac{\left(W_{4} \boldsymbol{u}_{i}\right)^{\top}\left(W_{5} \boldsymbol{u}_{k,i}+W_{3} \boldsymbol{f}_{ik}\right)}{\sqrt{D_h}}\right), 
\end{align}
\end{subequations}
where $W_1$, $W_2$, $W_3$, $W_4$, $W_5$ are trainable weights for one $\mathcal{T}$ operator; $N_i$ is the set of neighbors of vertex $i$, $N_i=\{k\in V:A(i,k)\neq 0\}$; $\boldsymbol{f}_{ik}$ is the edge feature of the edge $(i,k)$; $\beta_{i,k}\in(0,1)$ is the attention coefficient \cite{vaswani2017attention} that measures the coupling strength between node $i$ and $k$. $\boldsymbol{u}_{k, i}$ is the \textit{passed} input vector of node $k$; it is modified from $\boldsymbol{u}_{k}$. Recall for all vertices, the first three elements of $\boldsymbol{u}_{k}$ are its absolute coordinates $\boldsymbol{x}_k=(x_k, y_k, z_k)$. We replace $\boldsymbol{x}_k$ with $\boldsymbol{x}_{k,i}$, the relative coordinates with respect to node $i$, and account for the periodic boundary conditions:
\begin{equation}
\label{eqn:period}
\begin{aligned}
& \boldsymbol{x}_{k,i} = \boldsymbol{x}_k - \boldsymbol{x}_i - \text{nint}(\boldsymbol{x}_k-\boldsymbol{x}_i),  
\end{aligned}
\end{equation}
where $\text{nint}$ is the nearest integer function. For example, if $x_k=0.8$ and $x_i=0.1$, $\text{nint}(x_k-x_i)$ gives 1 and $x_{k,i}=-0.3$. Thus, the couplings of nodes are only sensitive to their distance and thus are translation invariant. If we translate the coordinate system, the coupling term won't change. In \Cref{sec:outofdis}, we will see this treatment allows GrainGNN to scale to large systems.

{\bf Regression decoder.}
The hidden states $H_l$ of the \textit{last} LSTM layer are passed to decoders. The decoder of model $\mathcal{R}$ is just a single layer perceptron: a linear layer with an activation function:
\begin{subequations}
\label{eqn:r_decoder}
\begin{align}
& \Delta x_{j}= \mathrm{tanh}(W_{hx}\boldsymbol{h}_j + b_x), \\
& \Delta y_{j} = \mathrm{tanh}(W_{hy}\boldsymbol{h}_j + b_y), \ \text{for } j  =1,..., n_j,\\
& \Delta s_g = \mathrm{tanh}\left( W_{hs} \boldsymbol{h}_{g} + b_s \right), \label{eqn:s_evo} \\
& {v}_{g,l} = \mathrm{ReLU}\left(W_{hv} \boldsymbol{h}_{g} + b_v\right)\label{eqn:v_evo}, \ \text{for } g  =1,..., n_g.
\end{align}
\end{subequations}
$W_{hx}, W_{hy}, W_{hs}, W_{hv}$ and $b_{x}, b_{y}, b_{s}, b_{v}$ are trainable weights and biases. Here $\Delta x_j, \Delta y_j, \Delta s_g, v_{g,l}$ are normalized outputs. We use  $\mathrm{tanh}$ activation for $\Delta x_j$, $\Delta y_j$, and $\Delta s_g$ because their values are in [-1, 1]. We use Rectified Linear Unit (ReLU) as the activation function for $v_{g}$ to ensure its non-negativity. From \Cref{eqn:s_evo} we compute the updated cross-sectional area $s_{g, l} = s_{g, l-1} + \Delta s_g$. Grain $g$ is removed from graph when $s_{g,l}<\epsilon_G$, where $\epsilon_G$ is a hyperparameter, that is determined by parameter sweep during network training. Typically $\epsilon_G=10^{-4}$ in our experiments. 

{\bf Regression loss function.}
Let $\Delta F$ be the training data and $\Delta \tilde{F}$ be the corresponding network prediction, the $L_2$-loss function for  $\mathcal{R}$ is:
\begin{equation}
\label{eqn:l2_loss}
L_2 =  \frac{1}{n_j} \sum_{j=1}^{n_j}\left[ \left(\Delta {x_j}- \Delta \tilde{x}_j \right)^2 + \left(\Delta {y_j}- \Delta \tilde{y}_j \right)^2 \right] + \frac{1}{n_g} \sum_{g=1}^{n_g} \left[ \left(\Delta s_{g}-\Delta \tilde{s}_{g} \right)^2 + \left(v_{g}-\tilde{v}_{g} \right)^2 \right].
\end{equation}

{\bf Classification decoder.}
The decoder of the classifier $\mathcal{C}$ predicts the probability of the edge event $P$ for each $E_{jj}$ edge. As shown in \Cref{fig:4graingnn}, $\mathcal{C}$ forms a vector that concatenates $\boldsymbol{h}_i$, $\boldsymbol{h}_j$, and $\boldsymbol{f}_{ij}$. Then $\mathcal{C}$ uses a linear layer followed by a sigmoid function to output the probability $P_{ij}\in (0,1)$: 
\begin{equation}
P_{ij} = \sigma \left( W_{hc}[\boldsymbol{h}_i, \boldsymbol{h}_j, \boldsymbol{f}_{ij}] + b_c \right), \ \forall (i,j)\in E_{jj},  
\label{eqn:c_decoder}
\end{equation}
where $W_{hc}$ and $b_c$ are trainable weights and biases. An $E_{jj}$ edge with probability $P_{ij}$ higher than $\epsilon_E \in (0, 1)$ is classified as a positive event of neighbor switching, where $\epsilon_E$ is the edge classification threshold.

{\bf Classification loss function.}
We use a binary cross-entropy (BCE) loss for $\mathcal{C}$. Let $Y_{ij}=\{0,1\}$ be the truth labels of whether $(i,j)$ is eliminated, the loss function is:
\begin{equation}
\label{eqn:bce_loss}
\begin{aligned}
& L_{CE} =  \frac{1}{|E_{jj}|} \sum_{\forall (i, j) \in E_{jj }} -Y_{ij}\text{log}P_{ij}-(1-Y_{ij}) \text{log}\left(1-P_{ij}\right).
\end{aligned}
\end{equation}

{\bf Network evaluation metrics.}
We use several metrics to evaluate the accuracy of the two networks.
For the regression network, we compute the Relative Root Mean Square Error (RRMSE) of the outputs: 
\begin{equation}
\text { RRMSE }=\sqrt{\frac{\sum_{i=1}^{n}\left(x_i-\tilde{x}_i\right)^{2}}{\sum_{i=1}^{n} x_i^2}} \times 100,
\end{equation}
where $x_i$ and $\tilde{x}_i$ are the ground truth and prediction respectively; $n$ is the number of samples. For classification accuracy, we use $F_1$-score.  $F_1$-score is the harmonic mean of the prediction precision and recall, where precision = TruePositive / (TruePositive + FalsePositive) and recall = TruePositive / (TruePositive + FalseNegative). Precision and recall both depend on the classification thresholds.  We calculate the area under the curve (AUC) of the precision-recall curve drawn from different thresholds; a higher AUC represents a more accurate classifier. From the precision-recall curves, we find the optimal classification thresholds $\epsilon_E$ and $\epsilon_G$ leading to the highest $F_1$-scores. The details of the training setup are provided in \Cref{sec:training}.

\subsection{Graph reconstruction and GrainGNN algorithm}
\label{sec:5graph_update}

Given the trained LSTM networks and classification thresholds, we apply the graph-to-graph update algorithm \Cref{alg:graingnn} that implements $(G_{l-1}, F_{l-1}) \rightarrow (G_l, F_l)$. This in essence the GrainGNN surrogate along with the image-to-graph and graph-to-image pre- and post-processing steps.

In GrainGNN's lines 2-3, the LSTMs $\mathcal{R}$ and $\mathcal{C}$ compute  $\Delta F$ and $P$ (\Cref{eqn:r_decoder}, \Cref{eqn:c_decoder}) respectively using the input features $F_{l-1}$. In lines 4-7, we update the features $x_j$, $y_j$, $z_l$, $s_g$, $v_g$ with $\Delta F$ and create the lists of events $\mathcal{S}_E$ and $\mathcal{S}_G$. Next, we update the graph with $\mathcal{S}_E$ and $\mathcal{S}_G$.

Note that during one graph update, some junctions are involved in more than one event for example $j_2$ in \Cref{fig:4update}a. Different orders of applying  $\mathcal{O}_E(j_1, j_2)$ and $\mathcal{O}_E(j_2, j_3)$ will result in different graphs. We address this by imposing an operator event ordering based on the predicted features. We next discuss this for grain elimination events and grain neighbor switching events.

{\bf Ordering grain elimination events}. We sort $\mathcal{S}_G$ by the predicted grain area $s_g$. Grains with smaller areas are expected to be eliminated first. Lines 11-20, define required graph topology updates when grain $g$ is eliminated. Recall a grain elimination event requires $|N_g|-2$ $O_E$ operations followed by an $O_G$ operation. We denote the $O_E$ operations of grain 
$g$ by  $\mathcal{S}_{E,g}$. An $O_E$ operation includes the updates of the edges as shown in \Cref{fig:3events}a. If $(j_1, j_2)$ is given as an edge event, we find their neighboring nodes $g_1,g_2,g_3,g_4,j_3,j_4,j_5,j_6$ by searching the edge list $E$ for edges having node $j_1$ or $j_2$. Then we perform edge replacements $(j_1, g_4)\rightarrow(j^{\prime}_1,g_2), (j_2, g_3)\rightarrow(j^{\prime}_2,g_1), (j_1, j_5)\rightarrow(j^{\prime}_2,j_5), (j_2, j_4)\rightarrow(j^{\prime}_1,j_4)$. $O_E$ also predicts the coordinates of $j^{\prime}_1$, $j^{\prime}_2$.  Currently, we do not implement this part in the networks; we approximate the coordinates of $j^{\prime}_1$ and $j^{\prime}_2$ simply with the midpoint of the edge $(j_1, j_2)$, i.e., $x_{j^{\prime}_1}=x_{j^{\prime}_2}=0.5(x_{j_1}+x_{j_2})$ (\Cref{fig:4update}a). We use this approximation due to the lack of training edge events, which is only about 2\% of the number of $E_{jj}$ edges.

{\bf Ordering grain switching neighbor events.} We need to perform this for both $\mathcal{S}_E$ and for switch events $\mathcal{S}_{E,g}$ triggered by grain eliminations. We discuss the latter first. 
In line 12 we initialize $\mathcal{S}_{E,g}$ as the edges of grain $g$. Then, we sort $\mathcal{S}_{E,g}$ by the area change of $g$'s neighboring grains $g_{ij}$, where $g_{ij}$ is the grain which shares the edge $(i,j)$ with $g$. In one $O_E$ operation, both $g$ and $g_{ij}$ lose edge $(i,j)$. Generally, a grain with an expanding cross-sectional area $\Delta s$ is more likely to gain faces rather than lose faces. Thus we choose to first update the edge of the grain whose $\Delta s$ is the smallest. For example in \Cref{fig:4update}b, the regressor $\mathcal{R}$ predicts $\Delta s_{g_{15}}<\Delta s_{g_{12}}<\Delta s_{g_{23}}<\Delta s_{g_{45}}<\Delta s_{g_{34}}$, so our updating order is $(j_1,j_5), (j_1,j_2), (j_2,j_3)$. In lines 16-19, we apply $O_E$ on each edge $(i,j)$ of $\mathcal{S}_{E,g}$ and if $(i,j)$ also appears in $\mathcal{S}_E$, we remove it so that it won't be updated again. In lines 21-24, we sort $\mathcal{S}_{E}$ by the elimination probability $P$ and apply $O_E$ on each event. For example in \Cref{fig:4update}a, $P_{12}>P_{23}$ so $(j_1,j_2)$ is updated first. We want to emphasize that although $j^{\prime}_1, j^{\prime}_2, j^{\prime}_3$ are temporarily at the same location, they are three different points and one of each only connects to three other junctions. Finally, in lines 25-28, we check the graph if it has 2-side grains, whose faces were eliminated during other events. We remove these grains although their area is still larger than 0.

{\bf Overall GrainGNN computational complexity.}
We analyze the complexity of \Cref{alg:graingnn}. The dominant cost of networks $\mathcal{R}$ and $\mathcal{C}$ is the evaluation of graph transformers. In \Cref{eqn:GT}, the cost of one matrix-vector multiplication is $O(D_h^2)$; therefore, the cost of one network inference of the entire graph is $O(n_gD_h^2)$. The cost of lines 4-7 is $O(n_g)$. For a graph update, the number of events $|\mathcal{S}_G|\sim O(n_g)$, $|\mathcal{S}_E|\sim O(n_g)$. Sorting $\mathcal{S}_E$/$\mathcal{S}_G$ in lines 10 and 22 is $O(n_g\text{log}(n_g))$ cost. One $O_E$ or $O_G$ consists of a couple of vertex and edge searches. Each event is $O(n_g)$ complexity. Thus the complexity per graph update is $O(D^2_hn_g+n^2_g)$. Recall that the time complexity $n_l$ is $O(1)$ so the overall complexity is $O(D^2_hn_g+n^2_g)$. In our experiments typically we have $n_g<D^2_h$ and the cost of GrainGNN roughly scales linearly with the number of grains $n_g$. 

\begin{algorithm}[h]
\small
\caption{GrainGNN graph-to-graph update algorithm $G_{l-1}(V_{l-1}, E_{l-1}), F_{l-1} \rightarrow G_l(V_l, E_l), F_l$\\
Parameters: $\epsilon_E$ and $\epsilon_G$ are classification thresholds}
\label{alg:graingnn}
\begin{algorithmic}[1]
\State  {\color{gray}/*  GrainGNN components $\mathcal{R}$ and $\mathcal{C}$ make predictions */ }
\State $\Delta F = \mathcal{R}(F_{l-1}, E_{l-1})$   \hfill {\color{gray} feature changes}
\State $P = \mathcal{C}(F_{l-1}, E_{l-1})$  \hfill {\color{gray} edge event probability}
\State $F_l \leftarrow F_{l-1} + \Delta F$ 
\State  {\color{gray} /* find events based on thresholds $\epsilon_E$, $\epsilon_G$ */ } 
\State $\mathcal{S}_E \leftarrow \{(m,n)\in E_{jj,l-1}:P_{mn}>\epsilon_E\}$   \hfill {\color{gray} edge events}
\State $\mathcal{S}_G \leftarrow \{g\in V_g:s_{g,l}<\epsilon_G\}$  \hfill {\color{gray} grain elimination events}
\State  {\color{gray} /* update graph with $\mathcal{S}_G$ */ } 
\State $V_l, E_l \leftarrow V_{l-1}, E_{l-1}$
\State $\mathcal{S}_G \leftarrow$ Sort $\mathcal{S}_G$ by $s_{g,l}$ in ascending order
\For { $g \in \mathcal{S}_G$}
    \State $\mathcal{S}_{E,g} = \{ (m, n) \in E_{jj,l}: m, n \in N_g \}$ 
    \hfill {\color{gray} edge events of eliminated grain} 
    \State $\mathcal{S}_{E,g} \leftarrow$ Sort $\mathcal{S}_{E,g}$ by $\Delta s_{g_{ij}}$ in ascending order
    \State $\mathcal{S}_{E,g} \leftarrow$ Remove last two elements
    \State  {\color{gray} /* update graph with $\mathcal{S}_{E,g}$ */ } 
    \For {$(i,j) \in \mathcal{S}_{E,g}$}
        \State $E_l, F_l \leftarrow O_E(E_l, F_l, i,j)$  \hfill {\color{gray} coordinates of new junctions} 
        \If{$(i,j)\in \mathcal{S}_E$}
            \State $\mathcal{S}_E \leftarrow$ Remove $(i,j)$ \hfill {\color{gray} avoid updating again in $\mathcal{S}_E$} 
        \EndIf
    \EndFor
    \State $V_l, E_l \leftarrow O_G(V_l, E_l, g)$ 
\EndFor

\State  {\color{gray} /* update graph with $\mathcal{S}_E$ */ } 
\State $\mathcal{S}_E \leftarrow$ Sort $\mathcal{S}_E$ by $P_{ij}$ in descending order
\For {$(i,j) \in \mathcal{S}_E$}
    \State $E_l, F_l \leftarrow O_E(E_l, F_l, i,j)$   \hfill {\color{gray} coordinates of new junctions}
\EndFor

\State  {\color{gray} /* remove 2-side grains */ } 
\State $\mathcal{S}^{\prime}_G \leftarrow \{g\in V_g:|N_g|=2\}$ 
\For { $g \in \mathcal{S}^{\prime}_G$ }
    \State $V_l, E_l \leftarrow O_G(V_l, E_l, g)$ 
\EndFor
    
\end{algorithmic}
\end{algorithm}

\subsection{Graph-to-image microstructure reconstruction}\label{sec:6graphTOimage}
We conclude with the post-processing step that converts the output GrainGNN to a 3D grain orientation field or directly to quantities of interest. Given $G_0, F_0$~\Cref{alg:graingnn} computes the trajectory  $(\mathcal{G,F}):=\{G_l,F_l\}_{l=0}^{n_l}$. 
We remark that with $\mathcal{G,F}$ we can directly compute the quantities of interest. For instance, the volume of a grain $g$ is given as $\mathcal{V}_g=\Delta z \sum^{n_l-1}_{l=0}s_{g,l} + v_g$. Therefore size, aspect ratio, and orientation statistics can be readily computed.  If needed, we can also reconstruct the pointwise orientation   $\boldsymbol{\theta}(x,y,z)$. We first reconstruct slices $I_l$ of the $x-y$ plane at different $z_l$ positions.  For each grain $g \in G_l$, we find its junction neighbors $N_g$ and their coordinates. We draw a polygon using the junctions and set  $I_l(x,y)=g$ at all interior pixels---at any desired pixel resolution. We repeat it for $n_g$ grains.
To combine all the $I_l$ slices we use piecewise constant interpolation in $z$: we assume that each reconstructed $I_l$ has thickness $\Delta z$ and then stack them in $z$-direction to form the 3D grain index field $\mathcal{I}$. We currently neglect the excess volume part of a grain in reconstructed 3D images. The final orientation field $\boldsymbol{\theta}(x,y,z)= \boldsymbol{\theta}_g(\mathcal{I}(x,y,z))$, where $\boldsymbol{\theta}_g$ is the orientation of a grain with index $g$. 

\subsection{Discretization and numerical solution of the phase field  PDE.}\label{sec:forward}
We use a second-order finite difference discretization in space and a forward Euler time stepping in time. The phase field code is implemented with Compute Unified Device Architecture (CUDA) in C++. Every grain is associated with a phase field function so one simulation will require storage of $n_g$ phase field functions. To reduce the complexity, we adopt an active parameter tracking (APT) algorithm \cite{vedantam2006efficient} to reduce the number of phase field variables stored on each grid point to a constant number $P$. Each grid point stores the largest $P$ numbers of $\phi_{\alpha}$ with its index $\alpha$ and treats all the other phase field variables as -1. In each time step at each grid point, only $\alpha$ stored in the point itself and its six direct neighbors compute \Cref{eq:micro_jsi}, and the new $P$ largest $\phi_{\alpha}$ are found and stored. APT reduces the storage by a factor of $n_g/(2P)$. 
We find $P=5$ is sufficient for our simulation setup. We also utilize the moving-domain technique \cite{Badillo2006} to reduce the height of the computational domain \footnote{We track a domain with a height smaller than the actual domain height $L^{\prime}_z$. The domain is placed around the SLI and moves with the SLI. As the SLI moves one grid point in the $z$-direction, we add a new layer of liquid on the top of the domain and remove the bottom layer of the solidified part. The removed layer is stored and remains ``frozen" till the end of the simulation.}.  
The convergence results of mesh size $dx$, interface width $W_0$, the number of phase field variables, and the height of the moving domain are reported in the appendix. 

But why an explicit scheme?  First, most practitioners \cite{pinomaa2019process,CHADWICK2021116862} use explicit schemes so this comparison is the most informative.  Second, although there are situations in which linearly-implicit or implicit-explicit solvers are preferable, this is not usually the case because in several rapid solidification regimes either the PDEs are not stiff or the implicit solvers become overly diffusive \cite{qin2022dendrite}. Third, implicit solvers complicate the implementation of the active parameter tracking. 

%% file: results.tex
\begin{figure}[h]
 \centering
     \includegraphics[width=1\textwidth]{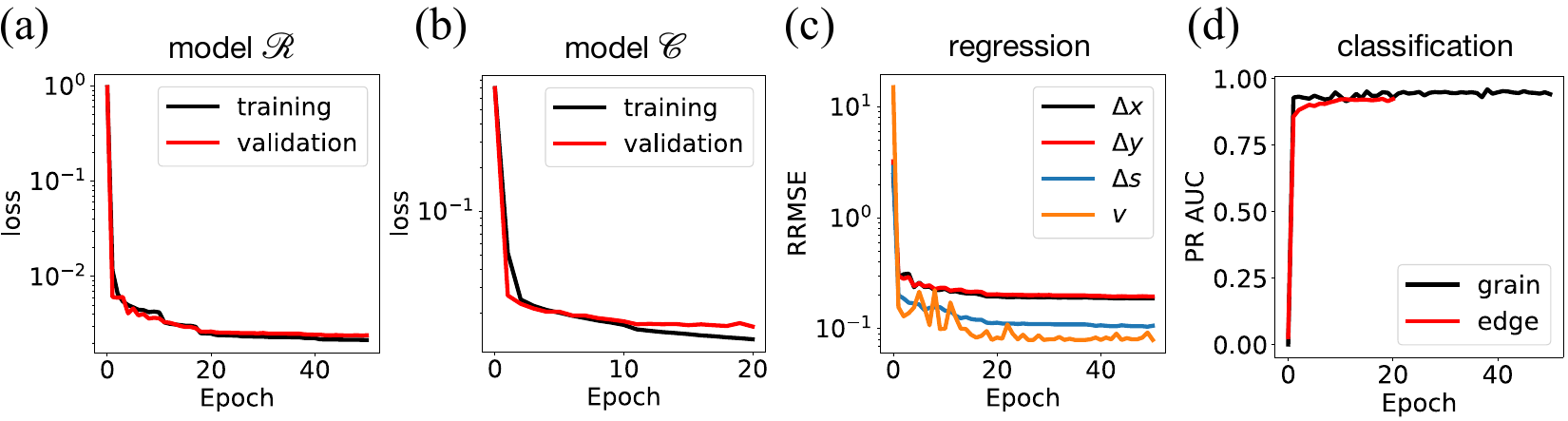}
     \caption{\textbf{Training losses and accuracy.} (a) Training and validation losses of the regressor $\mathcal{R}$. (b) Training and validation losses of the classifier $\mathcal{C}$. (c) Relative errors (RRMSE) of the regression outputs for validation data. $\Delta x, \Delta y, \Delta s$, and $v$. (d) The area under the Precision-Recall curve (PR AUC) of the predictions of the grain elimination events and neighbor-switching events. }
     \label{fig:5loss}
\end{figure}

In this section, we present training and testing experiments for GrainGNN. In \Cref{sec:training} we discuss the data generation for training, the network architecture parameters and number of parameters, metrics of comparison with simulations, and the training accuracy we obtained. In \Cref{sec:indis} and \Cref{sec:outofdis}, we discuss the testing accuracy. Recall that the input parameters to GrainGNN are $\piz=\{G_z,R_z\}$ (temperature profile), $\boldsymbol{\xi}$ the grain initial condition at $z_l=z_0$ (substrate of meltpool), and the domain dimensions $L=(L_x,L_y, L_z)$. For the temperature, we select a range of values for $G_z$ and $R_z$ and we select some values for training and different values for testing. The range of values corresponds to temperature profiles in metal alloy rapid solidification conditions.  For training $L$ is fixed to $L=L^0$ and $\boldsymbol{\xi}$ is sampled from a fixed distribution. For testing, we consider two consider two settings for $L$ and $\boldsymbol{\xi}$. (1) \emph{In distribution generalization} (\Cref{sec:indis})  we use $L=L^0$  and $\boldsymbol{\xi}$ is sampled from the same distribution used for training. (2) In our \emph{Out of distribution generalization} experiments (\Cref{sec:outofdis}),  we use $L\neq L^0$ and $\boldsymbol{\xi}$ is sampled from a different distribution used in training.  The latter is critical because although we train for a relatively small number of grains, we can generalize GrainGNN to an arbitrary number of grains and initial conditions without further training.

\subsection{Training of LSTM regressor and classifier}
\label{sec:training}
{\bf Selecting input parameter values for training.}
We use phase field data of different $\ksiz$ and $\piz$ as our training data. For each simulation, we use the same domain size $L^0$. We try to use as small $L^0$ as possible to minimize training costs.  We choose $L^0_x=L^0_y=40\mu m$  to have a sufficient initial number of grains to avoid the effect of periodic boundary conditions. We choose $L^0_z=50\mu m$ to have sufficient grain coarsening. The percentage of eliminated grains at $z_l=50\mu m$ is typically 50\%$-$70\%. For $\piz$, we used a uniform grid sampling with $G_z$ values in $(0.5, 10)\ K/ \mu m$ \cite{PINOMAA20201, yang2021phase,CHADWICK2021116862}, and $R_z$ values in $(0.2, 2)\ m/s$ \cite{CHADWICK2021116862,Pinomaa2020}. The sampling grid is shown in \Cref{fig:err_stat}a. The mesh size we use is $\Delta G=0.5K/\mu m$ and $\Delta R_z=0.2m/s$, thus the total number of sampled $\piz$ points is 1443. For each $\piz$ point we sample a different $\ksiz$. Grain orientations are uniformly sampled from a unit sphere, as discussed in \Cref{sec:pf}. The sampling of the initial junction coordinates $\boldsymbol{x}^0_j$ is as follows.


We first initialize grains as hexagonal lattices of the same size. Here we set their equivalent diameter $d_0=4.1\mu m$ \cite{xu2022three}. The junction coordinates of the hexagonal grid is $\boldsymbol{\bar{x}}_j$. Then we add a perturbation $\boldsymbol{x}^0_j=\boldsymbol{\bar{x}}_j + 0.1L^0_x\eta$ and $\eta$ is sampled from a Gaussian distribution $\eta \sim \mathcal{N}(\boldsymbol{0},\boldsymbol{I})$. For training data, the initial number of grains is in the range of 110$-$125 and the grain sizes are in the range of 2.7$-$5.7$\mu m$. For in-distribution generalization, testing $\ksiz$ is sampled using the same method. For out-of-distribution generalization, we sample $\boldsymbol{x}^0_j$ from a uniform distribution  $\boldsymbol{x}^0_j\sim U(0,1)$ to test grain sizes with a larger variance.  

{\bf Generating training data with phase field simulations.}
To generate training data we use high fidelity phase field simulations with the discretization described in \Cref{sec:forward}. We use the interface width $W_0=0.1\mu m$ and $dx=0.8 W_0$. We performed the convergence test to verify that this mesh size provides sufficient accuracy of quantities of interest for the $\piz$ range we used. The number of grid points used is $500\times500\times625$. 
The time step is 2.42 nanoseconds. The simulation is stopped when the SLI reaches $50\mu m$. The number of time steps varies between 10K for $R_z = 2m/s$ to 100K for $R_z = 0.2m/s$. 
The total cost for generating training data was about 15 hours on 24 NVIDIA A100 GPUs. 

From each simulation, we extract graph pairs $(G_{l-1}, G_l)$. For $z_l=L^0_z=50\mu m$, the number of eliminated grains varies between  0$-$84 so the number of extracted layers per simulation $n_l$ is 2$-$21. The total number of graph pairs created is 40K (38K for training and 2K for validation).  We first train the regressor network  $\mathcal{R}$  for 50 epochs with the Adam optimizer~\cite{zhang2018improved}.  The learning rate was set to 50\% decay every 10 epochs. Then we use $\mathcal{R}$'s weights to initialize the encoder of $\mathcal{C}$ which is trained for an additional 20 epochs again using the Adam optimizer.  The network hyperparameters---$D_h$, the number of LSTM layers, the batch size, and the initial learning rate---were tuned by grid search. The tuned $\mathcal{R,C}$ have two LSTM layers with $D_h=96$ for a total  1.2 million weights per network. The total training time was 12 hours on a single A100 GPU.  We remark that $\mathcal{C}$ and $\mathcal{R}$ run on both CPUs and GPUs. But during inference we also need operations $O_E$ and $O_G$, which we have implemented only on CPUs. So currently we run our inference only on CPUs and the inference can be further accelerated if we run $\mathcal{C}$ and $\mathcal{R}$ on GPUs. 


\Cref{fig:5loss} shows the training losses and accuracy of GrainGNN. For model $\mathcal{R}$, the initial training and validation losses are 0.97 and the final training and validation losses are 0.002. The corresponding accuracy of the output features is shown in \Cref{fig:5loss}c. At epoch 50, the RRMSE for $\Delta x, \Delta y, \Delta s$ and $v$ are 18.7\%,  19.5\%, 10.6\%, and 7.9\%, respectively. The average in-plane movement of a junction between two sampled $z_l$ values is roughly 2$-$3 pixels in $x$ and $y$ direction.
Thus, the error of the predicted junction coordinates is about half a pixel per graph update. 
Accuracy of grain events and edges events are shown as the black and red line in \Cref{fig:5loss}d. The final AUCs of grain events and edge events are 0.941 and 0.923 respectively. The optimal classification threshold for grain events we find on the precision-recall curves is $\epsilon_G=10^{-4}$; the corresponding precision and recall are 0.95 and 0.92. For edge events, we find $\epsilon_E=0.6$; precision is 0.91 and recall is 0.87.

Next we evaluate the accuracy of GrainGNN-predicted microstructure. $I_0$ is the PDE-to-image translation of $\ksiz$, the initial condition for both GrainGNN and phase field PDE. We extract $G_0, F_0$ from $I_0$ and compare the final predicted microstructure with phase field simulations. We use a pixel misclassification rate (MR) to measure pointwise errors.  MR$(z_l)$ is  as the number of grid points classified to a wrong grain, normalized by the total number of pixels:
\begin{equation}\label{eq:mr}
    \text{MR}(z_l) = \frac{1}{|I_l|}\sum_i\sum_j \mathbb{1}_{I_{l,\text{PF}}(i,j) \neq I_{l,\text{GNN}}(i,j)}.
\end{equation}
MR is zero when all grain boundaries are exactly reconstructed. For the quantities of interest, we measure the distributional error for grain sizes predicted by phase field simulations and GrainGNN. We use the  two-sample {Kolmogorov–Smirnov (KS)} statistics $\text{KS}=\sup _{x}\left|D_{\text{PF}}(x)-D_{\text{GNN}}(x)\right|$.  $D$ represents the empirical cumulative distribution function (eCDF) of grain sizes. KS statistics quantify the maximum discrepancy between the predicted and the ground truth grain size distributions. The smaller the KS value is the closer the two distributions are. More details about how to interpret KS statistics are provided in the appendix. 

\begin{figure}[h!]
 \centering
     \includegraphics[width=1\textwidth]{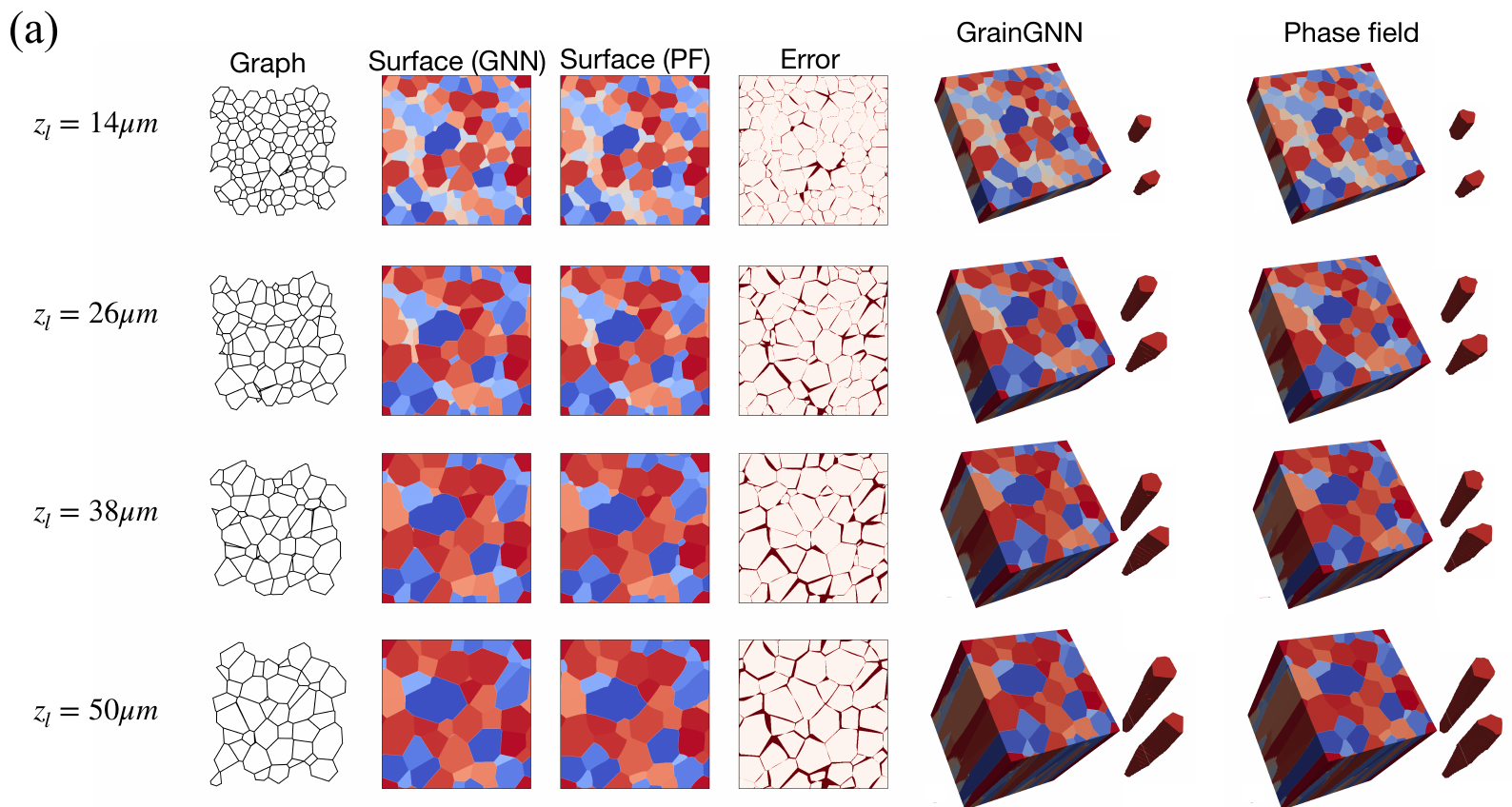}
     \includegraphics[width=1\textwidth]{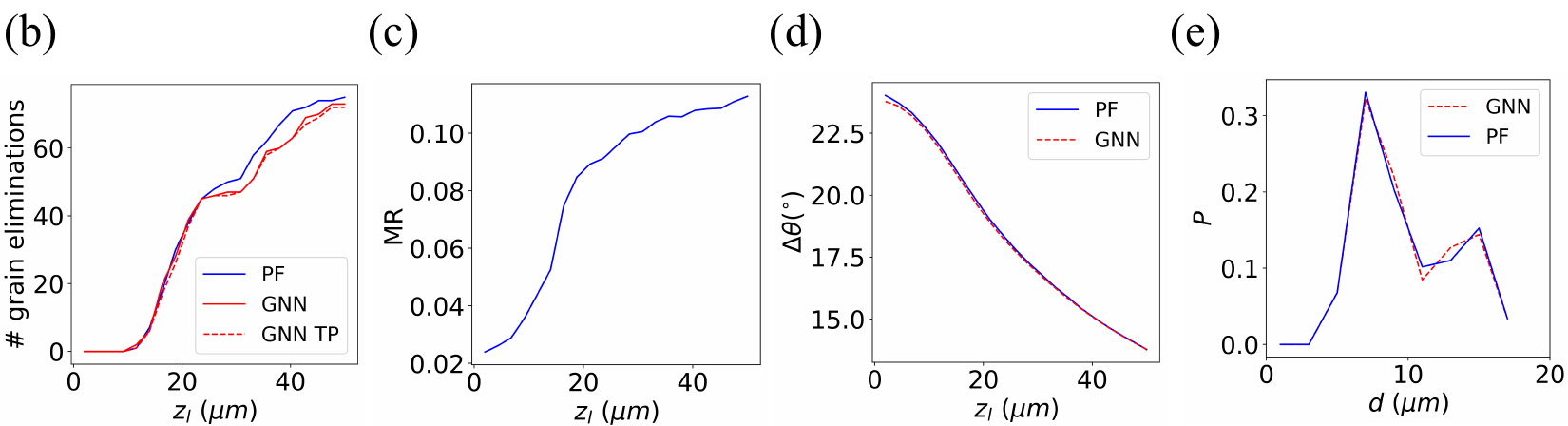}
     \caption{\textbf{GrainGNN prediction for one testing case with $G_z=1.904\ K/\mu m$ and $R_z=0.558\ m/s$.} (a) The time evolution of the grain microstructure when the solid-liquid interface $z_l$ reaches different heights. At each height from left to right, we show the graph that GrainGNN predicts, the reconstructed microstructure at the current height, the corresponding phase field simulation, the pointwise error between the GrainGNN image and the phase field result, the GrainGNN grain structures which are formed by stacking reconstructed images, and the phase field microstructures. We also explicitly show two of the grains to illustrate the evolution of the grain shape. (b) The number of accumulated grain elimination events when SLI reaches different heights. The solid blue line is the phase field result. The red line is predicted by GrainGNN. The red dashed line is the number of true positive (TP) events among the elimination events predicted by GrainGNN. (c) Misclassification rate (MR) at different heights. (d) Evolution of volume-average misorientation as the SLI reaches different heights. (e) The grain size distribution when $z_l=50\mu m$. The blue solid line and the red dashed line are phase field and GrainGNN predictions, respectively.}
     \label{fig:time_evo}
\end{figure} 

\subsection{In-distribution generalization}
\label{sec:indis}

By in-distribution generalization we refer to GrainGNN inference for $L=L^0$, unseen (during training) $\piz$ values, and unseen $\ksiz$ values sampled by the same procedure we used for training.  We show an example of test case in \Cref{fig:time_evo}  with $\piz=(1.904\ K/\mu m, 0.558\ m/s)$. Using nearest neighbor interpolation in $\piz$ space we determine that GrainGNN should be used with $\Delta z = 2.4\mu m$, which determines the number of GrainGNN steps. \Cref{fig:time_evo}a shows the grain microstructure at four different heights. At each height $z_l$, we show the $G_l$ and $I_l$ with  $500 \times 500$ spatial resolution. We compare the reconstructed image with the phase field data at the same height. The misclassified grid points are marked in dark red. The error images indicate that the GrainGNN predictions are accurate representations of the microstructure obtained by our phase field simulations. The last two columns compare the evolution of the 3D microstructure for GrainGNN and phase field predictions. We select two grains to illustrate the grain shape evolution. \Cref{fig:time_evo}b-e are different measures of GrainGNN accuracy. \Cref{fig:time_evo}b shows the cumulative grain elimination events when the SLI reaches different heights. The blue, red, and dashed red lines depict grain eliminations that happened in the phase field simulation only, GrainGNN inference only, and in both, respectively. We can see that the number of eliminations predicted by GrainGNN is overall accurate but lagged after $z_l=24\ \mu m$ compared to the phase field result. At $z_l=50\ \mu m$, three grains that should be eliminated still exist on the graph, and only one grain is falsely eliminated by GrainGNN. The precision and recall for grain events are 72/73 and 72/75 respectively. \Cref{fig:time_evo}c plots MR for reconstructed images at different heights. MR is 2.4\% initially at $z_0=2\ \mu m$ because the curvature of the grains is neglected in reconstructed images. MR is 11.3\% at $z_l=50\ \mu m$. \Cref{fig:time_evo}d shows the evolution of the volume-averaged misorientation as the height of the SLI increases; the curve is well captured by GrainGNN. The initial average misorientation angle is $23.8^{\circ}$ and it decreases to $13.8^{\circ}$, which indicates the grains are more aligned with the temperature gradients after the epitaxial growth. \Cref{fig:time_evo}e shows the grain size distribution when the SLI reaches $50\mu m $. The final average grain size is $9.9\mu m$. The KS statistic between the phase field and GrainGNN distributions is 0.034, which means the two distributions are nearly identical.

\begin{figure}
 \centering
     \includegraphics[width=1\textwidth]{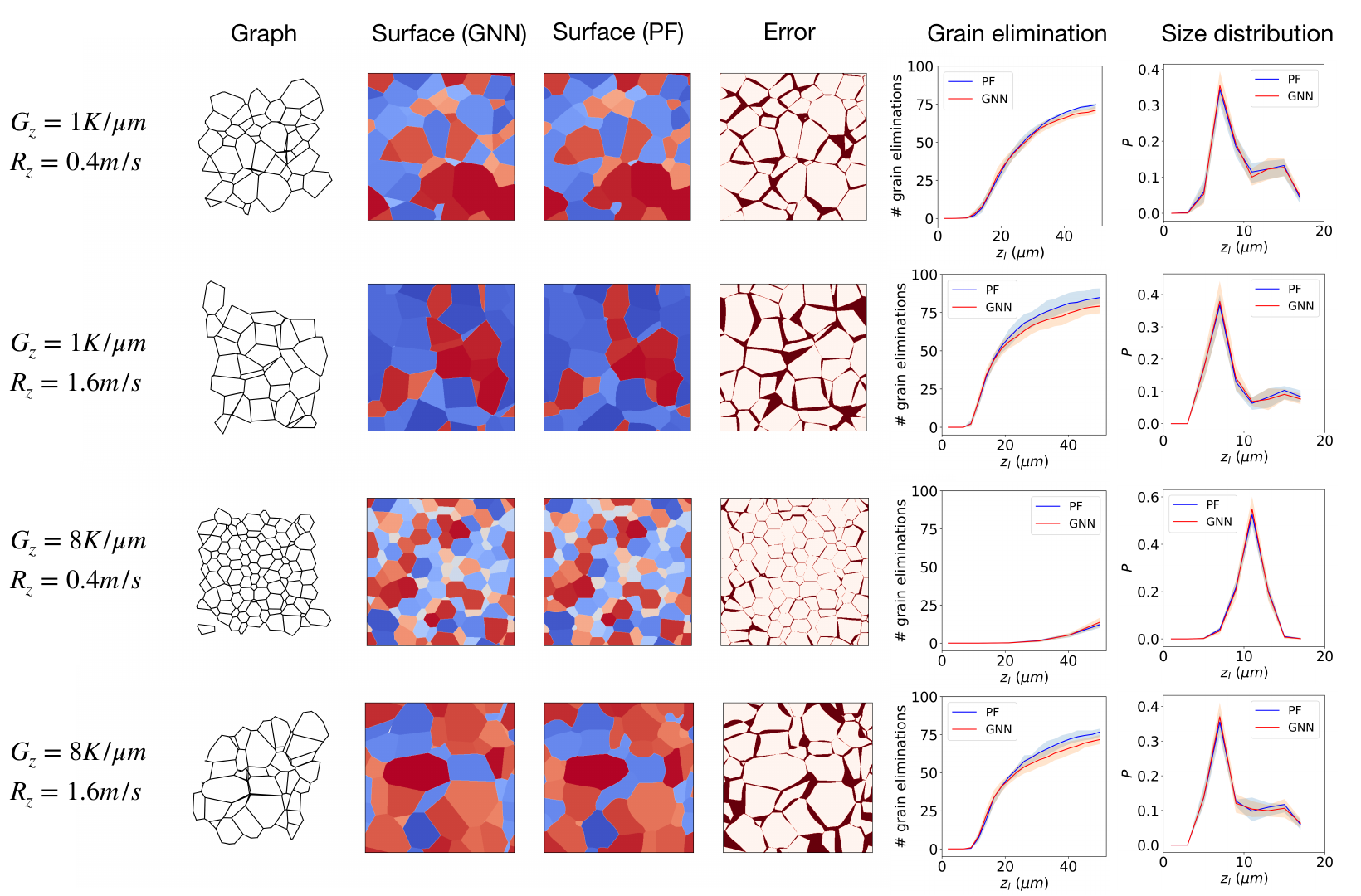}
     \caption{\textbf{GrainGNN predictions compared to phase field simulations of different $\piz$.} We run 10 $\ksiz$ for each $\piz$. The simulations are initialized with the same $\ksiz$ distribution. The 2D images show the grain structures at height $z_l=50\mu m$. Different $\piz$ result in different numbers of grain eliminations and final grain size distributions. Each blue/red line shows the mean and standard deviation of the 10 $\ksiz$ realizations.}
     \label{fig:GR}
\end{figure}

\begin{figure}
 \centering
     \includegraphics[width=1\textwidth]{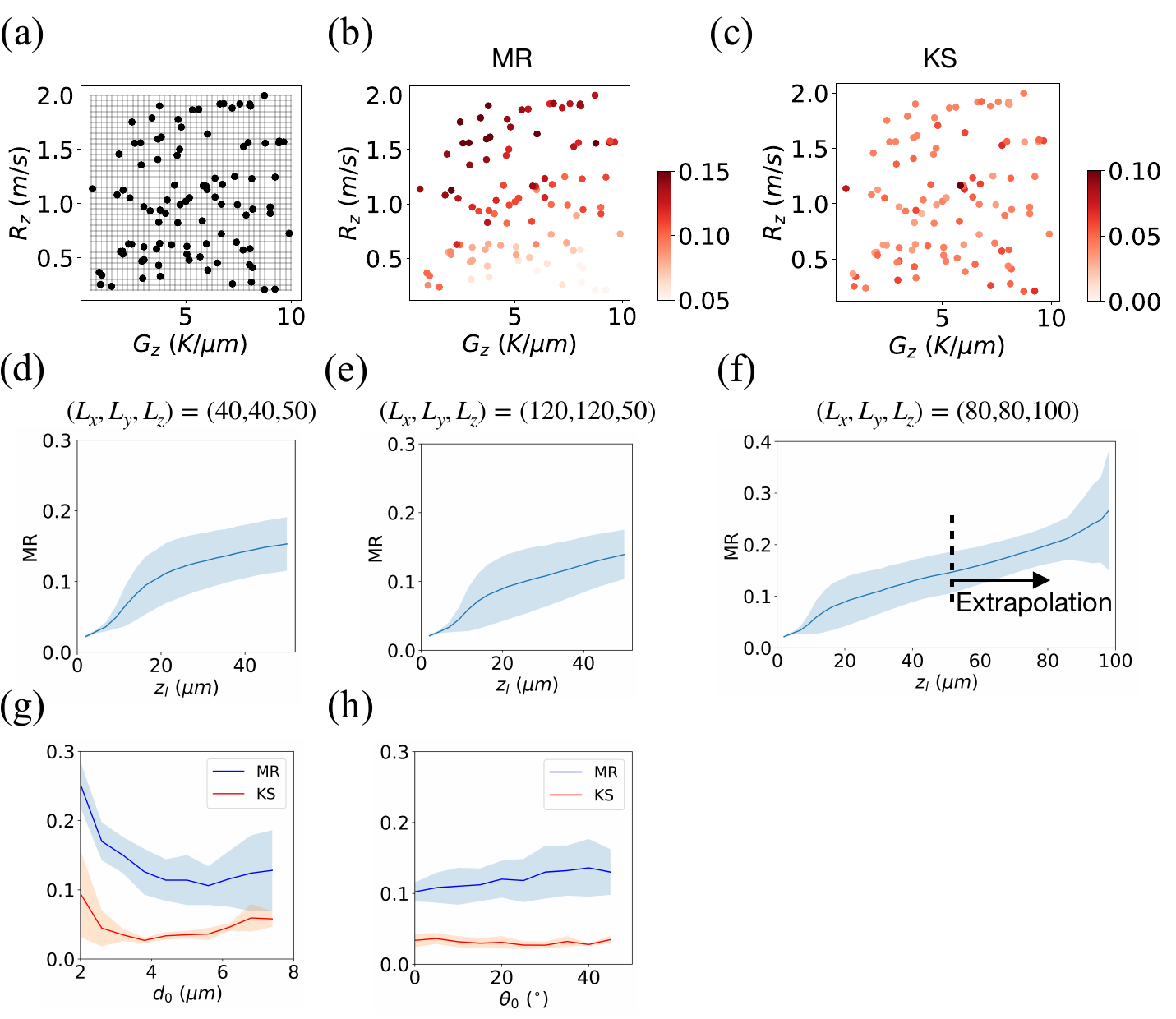}
     \caption{\textbf{Error statistics for in-distribution and out-of-distribution generalization.} (a) Black lines are $\piz$ grid used for training and validation. Black dots are 100 $\piz$ values for testing. (b) Misclassification rate (MR) averaged across sampled heights $z_l$ for each testing $\piz$. (c) KS statistics (KS) of grain size distribution at the end of simulations for testing $\piz$. (d) MR mean and standard deviation for 100 testing $\piz$ with domain size $(40\mu m, 40\mu m, 50\mu m)$. (e) MR for domain size $(120\mu m, 120\mu m , 50\mu m)$. (f) MR for domain size $(80\mu m , 80\mu m , 100\mu m)$. (g) MR and KS for different initial mean grain sizes $d_0$. (h) MR and KS for different initial grain orientation distributions. $\theta_0$ is the most frequent misorientation angle between the grain orientation and the $z$-axis. }
     \label{fig:err_stat}
\end{figure}  

We test the accuracy of GrainGNN inference for different values of  $\piz$. In our first test, we select four values of $\piz$, each of which we sample ten $\ksiz$ to compute microstructure statistics. In \Cref{fig:GR}, the 2D images show the grain microstructure at height $z_l=50\mu m$. We can see that lower $G_z$ and higher $R_z$ result in more grain eliminations and more misclassified pixels. The MR average and standard deviation at $z_l=50\mu m$ for the four values of $\piz$ are 16.2\%$\pm$1.7\%, 23.1\%$\pm$3.2\%, 8.6\%$\pm$0.5\%, and 18.8\%$\pm$2.3\%, respectively. The blue and red lines are the mean and standard deviation of the phase field and GrainGNN predictions. The KS average and standard deviation at $z_l=50\mu m$  are 4.7\%$\pm$0.7\%, 5.4\%$\pm$0.9\%, 5.2\%$\pm$0.9\%, and 4.4\%$\pm$0.6\% for the four $\piz$ values.

We further test GrainGNN's MR accuracy on 100 randomly selected $\piz$ with a single random $\ksiz$ for each value of $\piz$
(see \Cref{fig:err_stat}a ). \Cref{fig:err_stat}b show the MR averaged over $z_l$ for each testing case. MR increases with increasing  $R_z$ and decreasing  $G_z$; the highest MR is 17\% for $G_z=2.450K/\mu m, R_z=1.753m/s$ and $G_z=3.764K/\mu m, R_z=1.901m/s$. We didn't find a correlation between KS and $\piz$; the maximum KS is 0.11 for $G_z=5.826K/\mu m, R_z=1.165m/s$. \Cref{fig:err_stat}d shows MR as a function of the height $z_l$. At the height $z_l=50\mu m$, the mean and standard deviation are 15.3\%$\pm$3.8\% for 100 $\piz$. The total number of grain elimination events across the 100 cases is 6,597; the number of true positives predicted by GrainGNN is 6,088; the number of false positives is 124. 

\begin{figure}[h!]
 \centering
     \includegraphics[width=1\textwidth]{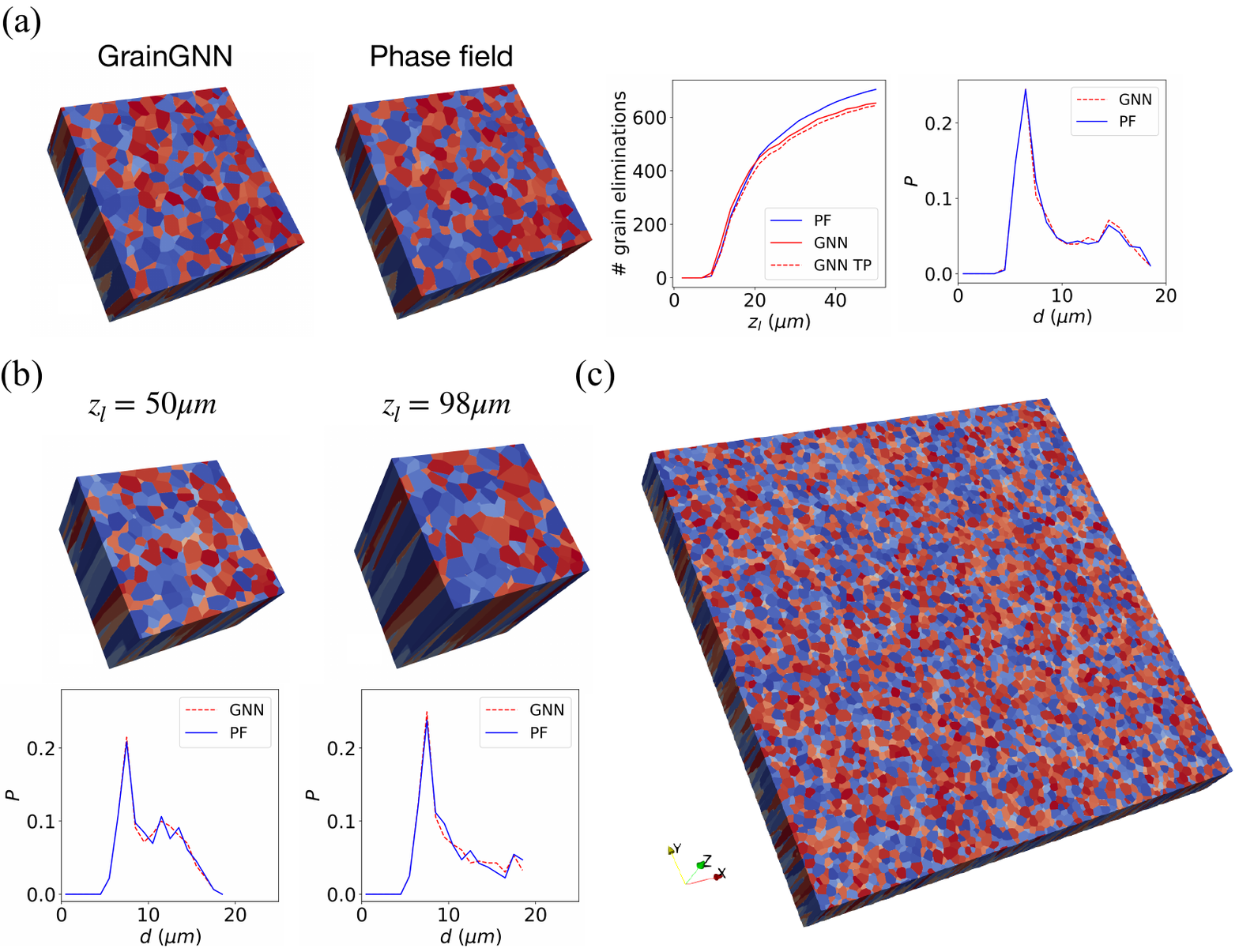}

     \caption{\textbf{Generalization with the domain size and the number of grains.} (a) Domain size ($120\mu m, 120\mu m, 50\mu m)$ with 1043 grains. $G_z=10K/\mu m, R_z=2m/s$. The total number of eliminated grains is 704, among which 644 grains are predicted by GrainGNN. The KS statistic for the grain size distribution is 0.021. (b) Domain size $(80\mu m, 80\mu m, 100\mu m)$ with 461 grains. $G_z=4.827K/\mu m, R_z=0.690m/s$. The left and right plots are GrainGNN predictions at $z_l=50\mu m$ and $z_l=98\mu m$, respectively. Their KS for grain size distribution are 0.024 and 0.028. (c) GrainGNN prediction for a case with domain size $(400\mu m, 400\mu m, 50\mu m)$, which is 100 times the training domain size. The total number of grains is 11600. $G_z=2K/\mu m, R_z=0.4m/s$. The GrainGNN inference time is 220 seconds on one CPU. }
     \label{fig:domain_extra}
\end{figure}

\subsection{Out-of-distribution generalization}
\label{sec:outofdis}

Recall that  $L^0=(40\mu m, 40\mu m, 50\mu m)$, mean substrate grain size $d_0=4.1\mu m$, and grain orientations from the unit sphere.  We examine GrainGNN's ability to predict 3D grain formation and QoIs for larger $L$ without changing the $\ksiz$ distribution. We also test GrainGNN's accuracy for different $\ksiz$ distributions. Note that these generalizations do not require any retraining of GrainGNN and only involve small modifications when $L>L^0$. We explain each generalization below.

\textbf{\textit{Domain width $L_x, L_y$}}: A larger domain width with the same grain size distribution is equivalent to a larger number of grains. As discussed in \Cref{sec:2graph}, the dimensional features in \Cref{eqn:grain_feature} and \Cref{eqn:junction_feature}, for example $x_j, y_j, s_g, v_g$ are normalized by the training domain size $L^0_x, L^0_y, L^0_z$. Although we change the domain size, we still normalize with $L^0$.
As shown in \Cref{fig:domain_extra}a, we have a testing case with domain width $L_x=L_y=120\mu m$. After normalization, the junction coordinates are in the range of [0, 3]. As the transformer encodes \textit{relative} distances
between a vertex and its neighbors, adding offsets to the coordinates won't change the hidden states output by the encoder.
Thus, we expect no difficulties in using a large domain.
We simply evolve GrainGNN on the entire graph and reconstruct the microstructure from the evolved graph. 

\Cref{fig:domain_extra}a shows an example of GrainGNN prediction for a case with domain size $(120\mu m, 120\mu m, 50\mu m)$. $G_z=10K/\mu m, R_z=2m/s$. The initial number of grains is 1043. At the end of the simulation, the number of eliminated grains is 704, of which 644 grains are predicted by GrainGNN. The MR at the top layer is 18.2\% and the KS statistic for the grain size distribution is 0.021. We further test 20 randomly sampled pairs of $(G, R)$ with domain size $(80\mu m, 80\mu m, 50\mu m)$, and repeat for domain size $(120\mu m, 120\mu m, 50\mu m)$. Their MR statistics are shown in \Cref{fig:err_stat}f and \Cref{fig:err_stat}e. At $z_l=50\mu m$, their average MR and standard deviation are 14.4\%$\pm$4.0\% and 13.9\%$\pm$3.6\%, respectively. Compared to 15.3\%$\pm$3.8\% for in-distribution generalization (\Cref{fig:err_stat}d), we don't observe a noticeable increase in pointwise error when increasing the domain width. The accuracy of accumulative grain events is 92.1\% (4,271/4,638) for $L_x=2L^0_x$ and 92.6\% (8,956/9,672) for $L_x=3L^0_x$, which is close to 92.3\% for $L_x=L^0_x$. In \Cref{fig:domain_extra}c, we showcase a GrainGNN inference with a large domain width $L_x=L_y=10L^0_x$. The total number of grains is 11,600. The number of vertex features of the graph is 313K. The final grain microstructure only requires 15 iterations of the graph, which takes about 220 seconds on one CPU. If we run a phase field simulation with the same configuration, the total number of grid points is $5000\times5000\times100=2.5$ billion. At each grid point if we store five phase field variables and their grain indices, the storage required is 25 billion numbers. The number of time steps for the phase field solver is  50K.

\textbf{\textit{Domain height $L_z$}}: In practice the domain height $L_z$ is determined by the meltpool depth. Here we test the ability of GrainGNN to generalize for $L_z>L^0_z$ by simply taking more iterations. Recall that we used  $L^0_z=50\mu m$. 
We assume for $z_l>L^0_z$, the graph evolution follows the same pattern as the previous graph step. In the network inference when $z_l>L^0_z$, we set $z_l$ as $L^0_z$ in \Cref{eqn:junction_feature} and \Cref{eqn:grain_feature}. 
As shown in \Cref{fig:err_stat}f, we extend the predictions of the 20 runs with width $L_x=2L^0_x$ to the height $z_l=98\mu m$. From $z_l=50\mu m$ to $z_l=98\mu m$, MR increases from 14.4\%$\pm$4.0\% to 26.6\%$\pm$11.6\% and the grain elimination accuracy drops from 92.1\% to  90.6\% (5,729/6,323). We can see from 50 to 90 $\mu m$ MR increases almost linearly with the distance the SLI traveled. From 90 to 98 $\mu m$, the jump of the MR is due to the drastic increase of error for a case with $G_z=1.774$ and $R_z=1.471$, whose MR reaches 71.5\% at $z=98 \mu m$.  \Cref{fig:domain_extra}b shows one case with $G_z=4.827K/\mu m, R_z=0.690m/s$. The KS of grain size distribution is 0.024 for $z_l=50\mu m$ and 0.028 for $z_l=98\mu m$.  

\begin{figure}[h!]
 \centering
     \includegraphics[width=1\textwidth]{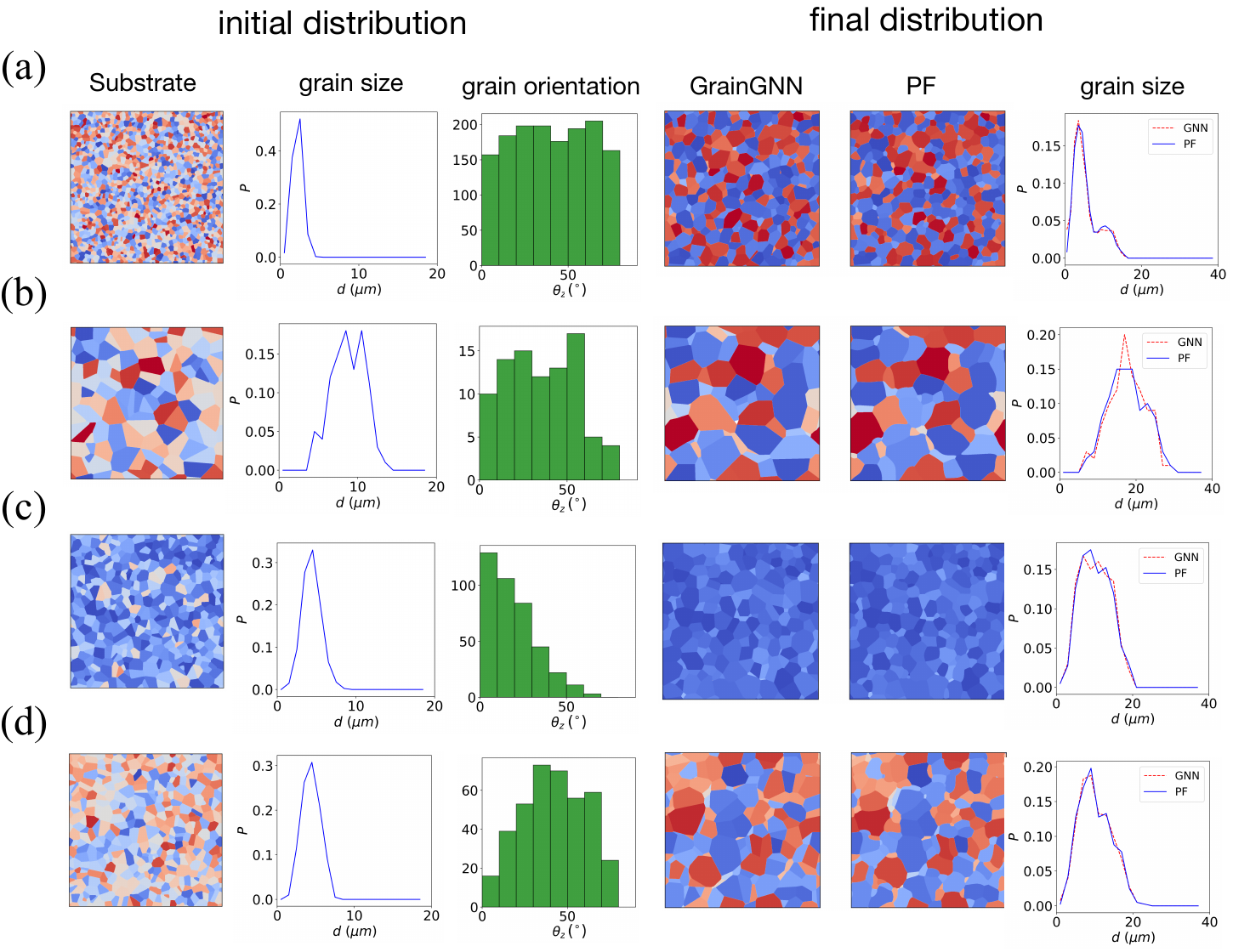}
     \caption{\textbf{Generalization with different initial grain configurations.} We use $L_x=L_y=80\mu m$, $G_z=4K/\mu m$, and $R_z=0.8m/s$ for all four cases shown above. (a) Initial size distribution with mean $d_0=2\mu m$. (b) Initial size distribution with mean $d_0=7.4\mu m$. (c) Initial orientation distribution with the maximum frequency at $\theta_0=0^{\circ}$. (d) Initial orientation distribution with the maximum frequency at $\theta_0=45^{\circ}$.}
     \label{fig:ini_extra}
\end{figure}  

\textbf{\textit{Initial grain size and orientation}}: We set $L=(80\mu m, 80\mu m, 50\mu m)\neq L^0$ and also vary $\ksiz$. As mentioned in \Cref{sec:training}, the initial junction coordinates are randomly selected from [0, 1]. We change the average grain size by varying the number of sampled junctions. 
\Cref{fig:ini_extra} shows the initial ($z=0$) and final ($z=L_z$)  grain distributions under $G_z=4K/\mu m$ and $R_z=0.8m/s$; we also test another four $\piz$ with the values in \Cref{fig:GR}.  For each $\piz$, we test ten different size distributions with mean ranging from $d_0=2\mu m$ (\Cref{fig:ini_extra}a) to $d_0=7.4\mu m$ (\Cref{fig:ini_extra}b). Their MR and KS are shown in \Cref{fig:err_stat}g where each point is averaged across five different $\piz$. For $d_0$ in the 4-6 $\mu m$ range, MR and KS are close to in-distribution errors. MR is higher for runs with smaller grains 
and is almost doubled for $d_0=2\mu m$. One reason is smaller grains have a higher surface area-to-volume ratio thus higher percentage of pixels will update their grain indices under the same physical parameters. Another reason is smaller $d_0$ has higher grain and edge event densities thus introducing larger errors. For larger grains $d_0>6\mu m$, we observe higher variance for different $\piz$. For $G_z=1K/\mu m$ and $R_z=1.6m/s$, MR increases when $d_0>6\mu m$ while MR decreases for the other four $\piz$. The overall MR for all $d_0$ is 13.9\% and grain elimination accuracy is 92.3\% (14,590/15,801). 

In the modified orientation distribution, we select dominant grain misorientation angle $\theta_0$ with respect to the $z$-axis. We vary $\theta_0$ from 0 to $\pi/4$ with ten values sampled for each $\piz$ and we sample the five $\piz$ as discussed for grain size distribution. \Cref{fig:ini_extra}c and \Cref{fig:ini_extra}d show the initial orientation distribution with $\theta_0=0$ and $\theta_0=\pi/4$ respectively. As shown in \Cref{fig:err_stat}h, MR is slightly higher for larger $\theta_0$, which is associated with more grain eliminations we observed for misaligned grains. KS is almost the same for different $\theta_0$. The overall MR and grain elimination accuracy are 12.0\% and 91.0\% (10,529/11,569), respectively.

%% file: discussion.tex
\subsection{Accuracy}

These results suggest that GrainGNN can  predict well the microstructure evolution and the statistics of quantities of interest when compared to the phase field simulations. It can also generalize to unseen $G_z$ and $R_z$ values,  domain width and height, and initial grain configurations---without retraining. For pointwise microstructure comparisons, the in-distribution generalization errors are in the 4\%$-$17\% range, with higher errors occurring when we have more grain eliminations. The out-of-distribution error is not sensitive to the domain width or the number of grains but increases with decreasing  initial grain size. The pointwise errors increase gradually with the height of the SLI. When comparing the statistics quantities of interest, we find the distributional error of grain size distribution is not sensitive to $G_z$, $R_z$, and the domain size. The KS statistics are in the 0.03--0.06 range. In all test cases, grain elimination accuracy is  above 90\%. In total, we ran 280 test phase field simulations with \textit{random} initial realization and the number of grains ranging from 100 to 1600. Demonstrating its robustness, GrainGNN successfully completed all 280 inferences, among which only one test case, with $L_z = 2 L^0_z$, yielded significantly different results.

We compare GrainGNN's accuracy with three other models, (i) a graph convolution LSTM (GC-LSTM) \cite{chen2022gc}, (ii) a graph transformer operator (TransformerConv) \cite{shi2020masked}, and (iii) a graph convolutional operator (GraphConv) \cite{morris2019weisfeiler}. We use the same training data and 2.4 million weights for all models. The metrics we used are RRMSE for $|\Delta \boldsymbol{x}| = \sqrt{\Delta x^2 + \Delta y^2}$ and $\Delta s$, AUC for both events, and the average MR for the 100 testing simulations. Compared to GC-LSTM, GrainGNN adds the attention coefficient $\beta_{i,k}$ in \Cref{eqn:GT}. As shown in \Cref{tab:models}, GrainGNN outperforms GC-LSTM for all the metrics. The self-attention mechanism of GrainGNN is important to learn the spatial correlations between different nodes. If we drop LSTM and only keep the TransformerConv as an encoder, regression accuracy RRMSE-$|\Delta \boldsymbol{x}|$ and RRMSE-$\Delta s$ decrease significantly. GraphConv has a higher AUC-G but lower RRMSE-$|\Delta \boldsymbol{x}|$, RRMSE-$\Delta s$, and AUC-E compared to TransformerConv. We also compute the average MR of 100 testing simulations for different models and GrainGNN produces the most accurate microstructure images.  

We also investigate the effect of the amount of training data on the GrainGNN accuracy. From \Cref{tab:models}, adding more training data generally improves the accuracy of GrainGNN. The regression model $\mathcal{R}$ is less sensitive to the size of training data. From 5K training pairs to 38K training pairs, the relative error of $|\Delta x|$ and $\Delta s$ improve 1\% every doubling the training size, and the AUC for grain elimination events stays around 94\%. In contrast, the AUC for the neighbor-switching events improves significantly from 85.2\% to 92.3\%. The difference between the two types of events is that the neighbor-switching events are highly imbalanced. The positive-to-negative ratio is about 1 to 30. Because the positive neighbor-switching events are deficient, adding more training graph pairs can largely enhance the classification precision. To deal with the data imbalance, we have tried to increase the weights of the positive events in \Cref{eqn:bce_loss} and downsampled the negative events; the improvement, however, is limited. We plan to use more data augmentation techniques to balance the labels and improve the accuracy of edge events.

The ordering of grain and edge events when updating the graph is another factor that affects the accuracy of the prediction. We currently choose the sorting method based on the classifier probabilities; we have not tested alternatives. One constraint of using \Cref{alg:graingnn} is the input graph can only have triple junctions and the output graph is guaranteed to have only triple junctions. \Cref{alg:graingnn} has two noteworthy limitations, although they do not seem to limit the overall GrainGNN accuracy. One is the approximation of new coordinates of $j_1, j_2$ in an edge event. The other is GrainGNN doesn't guarantee planar graph outputs, which means non-physical edge intersections are possible. The intersections could cause grains to have overlapping areas or create holes in the reconstructed images. We currently ignore this issue since it won't break \Cref{alg:graingnn} or calculations of quantities of interest.  

\begin{table}
\centering
\small
\caption{Model accuracy comparison for three network architectures and different numbers of training graph pairs. The total number of trainable parameters is roughly 2.4 million for all the models. For the regression tasks, we compute the Relative Root Mean Square Error (RRMSE) of $|\Delta \boldsymbol{x}|$ and $\Delta s$; for the classification tasks, we compare the area under the precision-recall curve (AUC) for neighbor-switching events (AUC-E) and grain elimination events (AUC-G). These metrics are evaluated on the same validation datasets. We also compute the average of misclassification rate (MR) for the 100 testing simulations with different $\piz$.}
\begin{tabular}{|c|cccccc|}
\hline
model & \#training pairs & RRMSE-$|\Delta \boldsymbol{x}|$ &RRMSE-$\Delta s$ & AUC-E & AUC-G & MR \\ \hline
\multirow{3}{*}{\specialcell{GrainGNN: LSTM + \\ TransformerConv }} & 5K & 0.225 & 0.143 & 0.852 & 0.938 & 0.140\\ 
 & 10K & 0.208 & 0.128 & 0.864 & 0.947 & 0.127\\
 & 20K & 0.197 & 0.119 & 0.892 & \textbf{0.953} & 0.121\\
 & 38K & \textbf{0.187} & \textbf{0.106} & \textbf{0.923} & 0.941  & \textbf{0.106}\\ \hline
LSTM + GraphConv \cite{chen2022gc} & 38K & 0.231 & 0.137 & 0.908 & 0.930 & 0.122\\ \hline
TransformerConv \cite{shi2020masked}  & 38K & 0.311 & 0.322 & 0.902  & 0.919 & 0.201\\ \hline
GraphConv \cite{morris2019weisfeiler}  & 38K & 0.388 & 0.394 & 0.847 & 0.936 & 0.250\\

\hline
\end{tabular}\label{tab:models}
\end{table}

\subsection{Computational efficiency}  
GrainGNN achieves a substantial reduction in the storage of variables and the number of required time steps, which leads to significant speedups over phase field simulations. For the training domain size, the required number of phase field variables is about 500M despite using the active parameter tracking and moving-domain algorithms, while the number of features of GrainGNN is about 3.2K. The network has about 2.4M parameters. Thus, GrainGNN requires $10^2-10^5$ times less storage than the phase field solver. The computation efficiency of our phase field solver and GrainGNN is listed in \Cref{tab:cost}. Our phase field solver is optimized with GPUs \cite{qin2022dendrite}. 
For the $G_z$ and $R_z$ values investigated in this paper, the time cost per phase field simulation is 300$-$3000 seconds on one A100 GPU, while the time cost of GrainGNN is 0.2$-$3 seconds. GrainGNN on a single CPU 
achieves 150$\times$--2000$\times$ speedup over our phase field code. 

The number of GrainGNN iterations $n_l$ scales linearly with the number of eliminated grains; therefore $n_l$ is a function of $G_z$ and $R_z$. In \Cref{tab:cost}, $n_l=5$ for $G_z=8\ K/\mu m$ and $R_z=0.4\ m/s$ and $n_l=20$ for other three parameters. We also compute the time consumed by network inferences with the domain width of $2L^0_x, 3L^0_x, 10L^0_x$. The number of grains per simulation is roughly 400, 1000, and 10,000, respectively. We can see the time cost scales linearly with the domain size and with the number of grains. If further increasing the number of grains to $n_g>10^4$, the  $O(n^2_g)$ graph update may start to dominate the cost. In this case, we plan to store a hash table of a node to its neighbors. Thus the cost per event can be reduced from $O(n_g)$ to $O(1)$. Another benefit of GrainGNN is that the cost doesn't increase with the grain size and number of grid points per grain. With increasing $L_z$ the graph size and inference time decrease due to grain eliminations. The inference time for $L_z=98\mu m$ is only approximately 1.5 times the inference time for $L_z=50\mu m$ (compared to approximately 2 times, if each height increment required a fixed amount of time).

\begin{table}
\centering
\small
\caption{Computational efficiency of GrainGNN for different $\piz$ and domain size $L$. Time for solving phase field equations was measured on one NVIDIA A100 GPU with 40 GB memory. The domain size is $(40\mu m, 40\mu m, 50\mu m)$. GrainGNN inference time was measured on a single AMD EPYC 7763 CPU core. $a_x\times a_y\times a_z$ represents the GrainGNN inference time for a domain size of $(a_xL^0_x, a_yL^0_y, a_zL^0_z)$. Every measured time is averaged across 10 initial realizations of grains.}
\begin{tabular}{|cc|cc|cccccc|}
\hline
\multicolumn{2}{|c}{Physical parameters} & \multicolumn{2}{|c}{storage} & \multicolumn{6}{|c|}{time (seconds)}\\ 
\hline
$G_z(K/\mu m)$ & $R(m/s)$ & PF & GrainGNN & PF & GrainGNN & $2\times2\times1$ & $3\times3\times1$ & $10\times10\times1$ & $2\times2\times2$ \\ 
\hline
1 & 0.4 & \multirow{4}{*}{500M} &\multirow{4}{*}{3.2K} & 1474.6 & 2.9 & 10.2 & 23.2 & 297.0  & 15.5\\
1 & 1.6 & & & 478.7  & 2.6 & 8.2 & 20.3  & 271.2  & 13.2\\
8 & 0.4 & & & 1440.9 & 0.9 & 3.5 & 7.8   & 69.5   & 6.4\\
8 & 1.6 & & & 413.4  & 2.8 & 9.6 & 21.8  & 278.7  & 15.1\\
\hline
\end{tabular}\label{tab:cost}
\end{table}


\subsection{Extensions and limitations}
There are several planned extensions to the current framework of GrainGNN. One is generalizing the computational domain to no-flux boundary conditions and non-rectangular geometries. For no-flux boundary conditions, we consider padding domain boundaries with halo grains that have mirrored properties (e.g., orientation) with respect to the grains on the boundary. The number of padding grains depends on the domain and physical parameters. For domain geometry, we plan to follow the rectangular-to-curvilinear domain mapping strategy presented in GrainNN \cite{qin2023grainnn}. The idea is to find geometric coefficients that map a curved surface to a 2D plane. We run GrainGNN in a rectangular domain and use the geometric coefficients to scale the network outputs for example the junction displacements. The output structure is mapped back to the original geometry. A second extension is to improve the representations of grain boundaries. We neglect the in-plane curvature of the grain boundaries, which can be significant for high $R_z$ values.
Another goal is to use experimental design and active learning \cite{lookman2019active,krogh1994neural} methods for sampling physical parameters used for training. Currently, we use a uniform grid of  $G_z$ and $R_z$ to generate training data. The MR plot in \Cref{fig:err_stat} indicates that more data should be drawn from the region with high $R_z$ and low $G_z$ to reduce the pointwise error. To make the data generation computationally trackable for higher dimensional parameter space, an efficient adaptive sampling algorithm needs to be developed. From the network architecture perspective, the current one-to-one LSTM prediction can be extended to sequence-to-sequence prediction to improve accuracy as we did in~\cite{qin2023grainnn}. However it is unclear how to handle topological changes in sequence-to-sequence configurations.

{\bf Limitations.} The two major limitations of GrainGNN are dealing with complex meltpool geometries and ignoring grain nucleation. We discussed geometry in the previous paragraph. Grain nucleation introduces new grains to the system thus adding vertices and edges to the graph. The modifications of the graph are two kinds depending on the nucleation density. For dilute nucleation, we can still utilize GrainGNN. We will introduce a new operation $O_N$ that adds vertices and edges, which will be the inverse operation of $O_G$. For dense nucleation, the equiaxed growth dominates the grain formation. This evolution is substantially different enough from epitaxial growth that a third network (beyond $\mathcal{R}$ and $\mathcal{C}$) will have to be added to predict the graph generation during grain nucleation.  A third more fundamental limitation is detecting failure cases, in short some kind of a posteriori error estimates for network predictions. GrainGNN doesn't have any performance guarantees other than the empirical evaluation we discussed. A fourth limitation is that the inference phase of GrainGNN runs only CPUs, and thus potential speedups are still possible. We are currently working to address these challenges.

%% file: conclusion.tex
We presented a surrogate for microstructure evolution in 3D epitaxial grain formation. We introduced a heterogeneous graph model and hand-crafted graph features that combined to achieve a significant spatiotemporal compression of grain microstructure. We proposed an image-to-graph method to extract graphs from phase field data. We modeled microstructure formation with graph-to-graph evolution, where we decomposed the evolution into feature changes and topological events. GrainGNN is implemented with an LSTM-based regressor and classifier to predict the features and events. The LSTMs include graph transformers and can scale to a larger number of grains than those used for training. We also proposed graph and microstructure reconstruction methods to address topological changes. GrainGNN is trained with phase field data in a wide range of $G$ and $R$ values in AM process conditions. It can predict both quantities of interest and pointwise accurate microstructure for unseen process parameters and initial grain configurations. We also showed that GrainGNN generalizes to domain size, number of grains, and initial grain parameters. GrainGNN is orders of magnitude faster than high fidelity simulations and scalable to a large number of grains.

What about training costs? Consider 3D printing a 1cm$^3$ volume part with an average $10^6\mu$m$^3$ meltpool. Directly simulating solidification of the entire part would require one million phase field meltpool solidification calculations---without accounting for ensemble calculations and meltpool overlaps. Such calculations are currently infeasible. A GPU-optimized GrainGNN on a multi-GPU leadership system could perform such a calculation in less than a day. Our long-term goal is to train GrainGNN using 1000s of phase field simulations using a small domain and then, by accounting for meltpool geometry, deploy it for inference for entire-part microstructure prediction. Thus, training costs will be amortized across the entire build and insignificant compared to the potential overall speedup.

%% file: acknowledgments.tex
This material is based upon work partially supported by NSF award OAC 2204226 and by the U.S. Department of Energy, Office of Science, Office of Advanced Scientific Computing Research, Applied Mathematics program, Mathematical Multifaceted Integrated Capability Centers (MMICCS) program, under award number DE-SC0023171. Any opinions, findings, and conclusions or recommendations expressed herein are those of the authors and do not necessarily reflect the views of the DOE, and NSF. Computing time on the Texas Advanced Computing Centers Stampede system was provided by an allocation from TACC and the NSF.

%% file: appendix.tex
\counterwithin{figure}{section}
\counterwithin{table}{section}

\section{Phase field solver}

The material parameters of stainless steel 316L are listed in \Cref{tab:comp_param}. Given values of $\Gamma$, $L_f/C_p$, and $\mu_k$, phase field parameters $\lambda$ and $\tau_0$ can be derived using asymptotic analysis \cite{bragard2002linking, qin2022dendrite}:
\begin{align}
    &\frac{\Gamma C_p}{L_f} \equiv d_0 = a_1\frac{W_0}{\lambda}, \\
    & \frac{C_p}{\mu_k L_f} \equiv \beta_0 = a_1\frac{\tau_0}{\lambda W_0} -a_1a_2\frac{W_0}{D_h} ,
\end{align}
where $a_1= 5\sqrt{2}/8$, $a_2 = 47/75$; $d_0$ is the thermal capillarity length; $\beta_0$ is the kinetic coefficient; $D_h$ is the heat diffusion coefficient. The obstacle parameter $\omega$ in \Cref{eq:micro_jsi} is set to $\lambda u/I$, where $u$ is the nondimensional undercooling and $I$ is a constant, here we choose $I=1/12$ \cite{ofori2010quantitative}. The anisotropy of $\tau_\alpha$ requires the spatial derivatives of $\phi$ in the rotated coordinates with angle $\boldsymbol{\theta}$. Let ${\theta}_z$ be the angle between the $z$-axis and the preferred growth direction <100> and ${\theta}_x$ be the angle between the $x$-axis and the projection of <100> direction on the $x$-$y$ plane. The derivatives appearing in \Cref{eqn:anis1} should be replaced with:
\begin{equation}
\left[\begin{array}{c} {\grad{}}_x\phi^{\prime} \\ {\grad{}}_y\phi^{\prime}  \\ {\grad{}}_z\phi^{\prime}  \end{array}\right] = \left[\begin{array}{ccc} \cos(\theta_x)\cos(\theta_z) & \sin(\theta_x)\cos(\theta_z) & -\sin(\theta_z) \\ -\sin(\theta_x)  & \cos(\theta_x) & 0 \\ \cos(\theta_x)\sin(\theta_z) & \sin(\theta_x)\sin(\theta_z) & \cos(\theta_z) \end{array}\right] \cdot\left[\begin{array}{c} {\grad{}}_x\phi \\ {\grad{}}_y\phi  \\ {\grad{}}_z\phi \end{array}\right]
\end{equation}
Because the capillary anisotropy is much weaker than kinetic anisotropy in the rapid solidification regime, we set $\epsilon_c=0$.

The phase field solver is available at https://github.com/YigongQin/cuPF. The CUDA functions have been optimized. The most expensive right-hand side calculations have the performance of 270 GFlops on one Nvidia A100 GPU.  We use CUDA-aware MPI which scales well on multiple GPUs and multiple nodes. 

\begin{figure}
 \centering
     \includegraphics[width=0.6\textwidth]{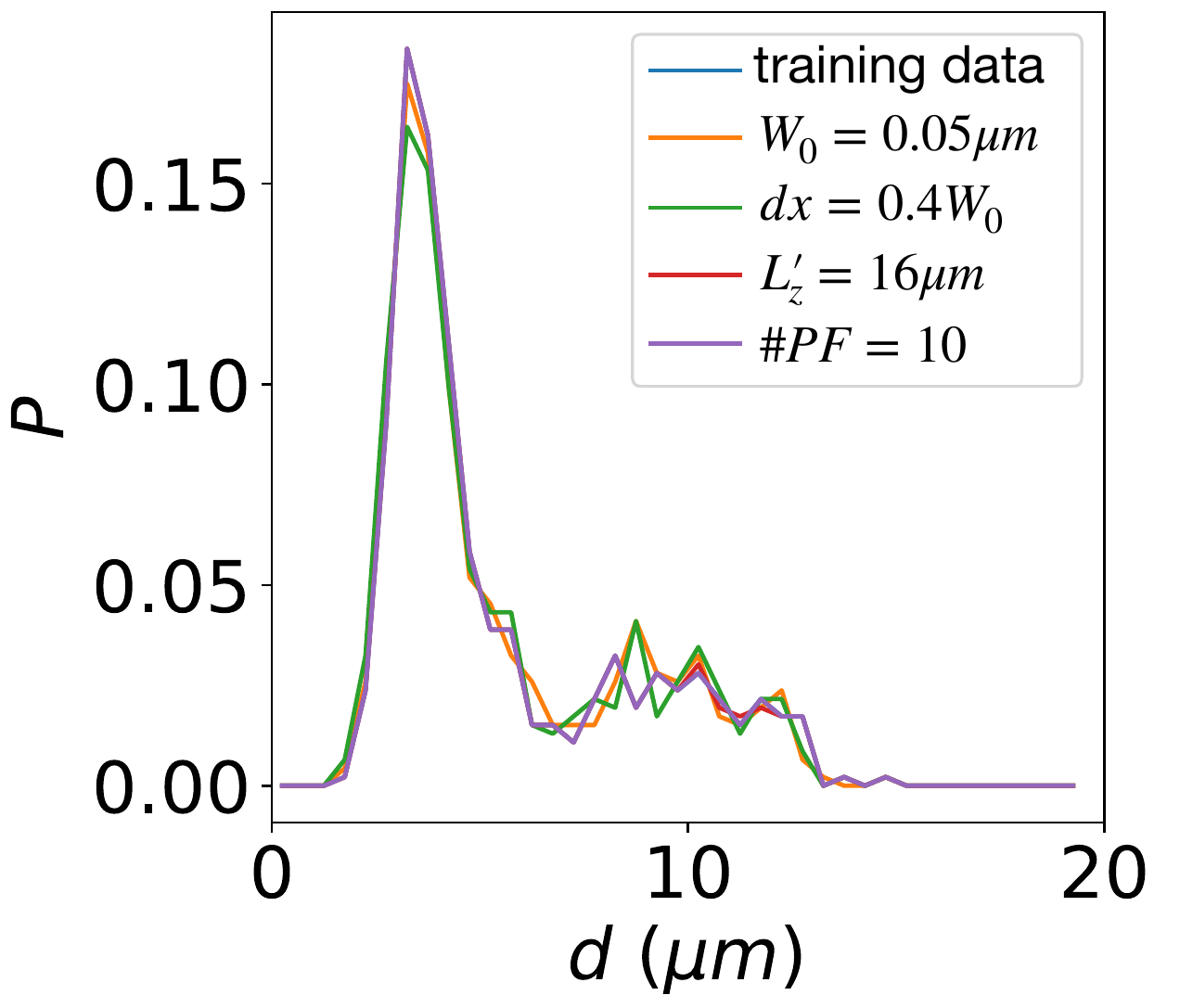}
     \caption{\textbf{Convergence test for a phase field simulation.} $G_z=10K/\mu m$ $R_z=2m/s$. The number of grains is 463. The simulation parameters used for generating the training data (blue line) are interface width $W_0=0.1\mu m$, mesh size $dx=0.8W_0$, moving-domain height $L^{\prime}_z=8\mu m$, and five phase field variables per grid point. The yellow and green lines show the convergence of the discretizations by halving $W_0$ or $dx$. The red line shows the convergence of the moving-domain algorithm by doubling the moving-domain width. The convergence of the blue and purple lines indicates the negligible difference between storing five phase fields and ten phase fields. }
     
     \label{fig:conv}
\end{figure}

\Cref{fig:conv} shows the convergence results for $dx$, $W_0$, the height of the moving domain, and the number of phase field variables per grid point used in the active parameter tracking algorithm. The training data is generated using the configuration: $W_0=0.1\mu m$, $dx=0.8W_0$, $L^{\prime}_z=8\mu m$, and five phase fields. We can see that the grain size distributions converge well when we halve the mesh size or increase the domain height or the number of phase fields.


\begin{table}
\centering
\caption{Material parameters for stainless steel 316L}
\begin{tabular}{|ccc|}
\hline
symbol & meaning (units) &  value \cite{CHADWICK2021116862,PINOMAA20201} \\
$\Gamma$ & Gibbs-Thompson coefficient (Km)  & $3.47\times 10^{-7}$\\
$L_f/C_p$  & latent heat/heat capacity (K)  & 229 \\
$\mu_k$  & linear kinetic coefficient (m/s/K)   & 0.217\\
$\lambda$ & thermal coupling constant  &  58.3\\
$W_0$ & length scale ($\mu m$)  & 0.1 \\
$\tau_0$ & time scale (ns) & 40\\
$T_M$ & melting temperature (K) &  1783 \\
$\epsilon_k$ & kinetic anisotropy coefficient  & 0.11 \\
k & partition coefficient  & 0.791\\
$D_h$ & heat diffusion coefficient ($m^2/s$)  & $3.6\times 10^{-6}$\\
$\Delta T_0$  & freezing range (K) &15.7 \\
$v_a$ & absolute stability velocity (m/s)  & 0.17 \\
\hline
\end{tabular}\label{tab:comp_param}
\end{table}

\section{Kolmogorov–Smirnov test}
 Kolmogorov-Smirnov test can be used to compare two samples and test whether the two samples could have come from the same distribution. If sample A has $m$ data points and sample B has $n$ data points, the null hypothesis that two samples coming from the same distribution is rejected when:
 \begin{equation}
     \text{KS}>\sqrt{-\ln \left(\frac{\alpha}{2}\right) \cdot \frac{1+\frac{n}{m}}{2 n}},
 \end{equation}
where KS is the Kolmogorov-Smirnov statistic and $\alpha$ is the confidence level. For our setup, if a testing case has $n=120$ grains and we think GrainGNN and phase field results are the same with a 95\% confidence level, the critical KS value is $\text{KS}_c = \sqrt{-\ln \left(\frac{0.95}{2}\right)/120}=0.079$. As discussed in \Cref{sec:indis}, the KS values for in-distribution generalization are mostly in the range of 0.03$-$0.06, meaning that GrainGNN predicts the same grain size distribution as the phase field model. For out-of-distribution generalization we change the number of grains. $\text{KS}_c$ is 0.043 for $n=400$ and 0.022 for $n=1600$. In \Cref{fig:err_stat}h we vary the initial orientation distribution with $n=400$. We can see $\text{KS} \approx \text{KS}_c$ for all $\theta_0$, which means we have 95\% confidence GrainGNN and phase field have the same grain size distributions. In \Cref{fig:err_stat}g when we decrease initial grain sizes, $\text{KS}>\text{KS}_c$ for $d_0<3\mu m$, which indicates a higher distribution mismatch for smaller grains.

\section{Code and data availability}
GrainGNN codes are available at https://github.com/YigongQin/GrainGraphNN. Codes contain how to create a graph using the Voronoi diagram and how to extract a graph from grain microstructure images. The neural networks are developed based on the Pytorch Geometric Library. Training and testing of the networks with comparisons against the phase field results are provided.  The GitHub repository also provides the trained models including the regressor and the classifier. The phase field data used for training can be reproduced with the CUDA codes at https://github.com/YigongQin/cuPF. 


%% file: main.bbl
\begin{thebibliography}{10}
\expandafter\ifx\csname url\endcsname\relax
  \def\url#1{\texttt{#1}}\fi
\expandafter\ifx\csname urlprefix\endcsname\relax\def\urlprefix{URL }\fi
\providecommand{\bibinfo}[2]{#2}
\providecommand{\eprint}[2][]{\url{#2}}

\bibitem{smith2016linking}
\bibinfo{author}{Smith, J.} \emph{et~al.}
\newblock \bibinfo{title}{Linking process, structure, property, and performance
  for metal-based additive manufacturing: computational approaches with
  experimental support}.
\newblock \emph{\bibinfo{journal}{Computational Mechanics}}
  \textbf{\bibinfo{volume}{57}}, \bibinfo{pages}{583--610}
  (\bibinfo{year}{2016}).

\bibitem{steinbach1999generalized}
\bibinfo{author}{Steinbach, I.} \& \bibinfo{author}{Pezzolla, F.}
\newblock \bibinfo{title}{{A generalized field method for multiphase
  transformations using interface fields}}.
\newblock \emph{\bibinfo{journal}{Physica D: Nonlinear Phenomena}}
  \textbf{\bibinfo{volume}{134}}, \bibinfo{pages}{385--393}
  (\bibinfo{year}{1999}).

\bibitem{ofori2010quantitative}
\bibinfo{author}{Ofori-Opoku, N.} \& \bibinfo{author}{Provatas, N.}
\newblock \bibinfo{title}{{A quantitative multi-phase field model of
  polycrystalline alloy solidification}}.
\newblock \emph{\bibinfo{journal}{Acta Materialia}}
  \textbf{\bibinfo{volume}{58}}, \bibinfo{pages}{2155--2164}
  (\bibinfo{year}{2010}).

\bibitem{PINOMAA20201}
\bibinfo{author}{Pinomaa, T.}, \bibinfo{author}{Lindroos, M.},
  \bibinfo{author}{Walbr{\"u}hl, M.}, \bibinfo{author}{Provatas, N.} \&
  \bibinfo{author}{Laukkanen, A.}
\newblock \bibinfo{title}{{The significance of spatial length scales and solute
  segregation in strengthening rapid solidification microstructures of 316L
  stainless steel}}.
\newblock \emph{\bibinfo{journal}{Acta Materialia}}
  \textbf{\bibinfo{volume}{184}}, \bibinfo{pages}{1--16}
  (\bibinfo{year}{2020}).

\bibitem{yang2021phase}
\bibinfo{author}{Yang, M.}, \bibinfo{author}{Wang, L.} \& \bibinfo{author}{Yan,
  W.}
\newblock \bibinfo{title}{Phase-field modeling of grain evolutions in additive
  manufacturing from nucleation, growth, to coarsening}.
\newblock \emph{\bibinfo{journal}{npj Computational Materials}}
  \textbf{\bibinfo{volume}{7}}, \bibinfo{pages}{1--12} (\bibinfo{year}{2021}).

\bibitem{CHADWICK2021116862}
\bibinfo{author}{Chadwick, A.~F.} \& \bibinfo{author}{Voorhees, P.~W.}
\newblock \bibinfo{title}{{The development of grain structure during additive
  manufacturing}}.
\newblock \emph{\bibinfo{journal}{Acta Materialia}}
  \textbf{\bibinfo{volume}{211}}, \bibinfo{pages}{116862}
  (\bibinfo{year}{2021}).

\bibitem{gandin19973d}
\bibinfo{author}{Gandin, C.-A.} \& \bibinfo{author}{Rappaz, M.}
\newblock \bibinfo{title}{{A 3D cellular automaton algorithm for the prediction
  of dendritic grain growth}}.
\newblock \emph{\bibinfo{journal}{Acta Materialia}}
  \textbf{\bibinfo{volume}{45}}, \bibinfo{pages}{2187--2195}
  (\bibinfo{year}{1997}).

\bibitem{rolchigo2022understanding}
\bibinfo{author}{Rolchigo, M.}, \bibinfo{author}{Carson, R.} \&
  \bibinfo{author}{Belak, J.}
\newblock \bibinfo{title}{{Understanding Uncertainty in Microstructure
  Evolution and Constitutive Properties in Additive Process Modeling}}.
\newblock \emph{\bibinfo{journal}{Metals}} \textbf{\bibinfo{volume}{12}},
  \bibinfo{pages}{324} (\bibinfo{year}{2022}).

\bibitem{rolchigo2022exaca}
\bibinfo{title}{{ExaCA: A performance portable exascale cellular automata
  application for alloy solidification modeling}, author={Rolchigo, Matt and
  Reeve, Samuel Temple and Stump, Benjamin and Knapp, Gerald L and Coleman,
  John and Plotkowski, Alex and Belak, James}}.
\newblock \emph{\bibinfo{journal}{Computational Materials Science}}
  \textbf{\bibinfo{volume}{214}}, \bibinfo{pages}{111692}
  (\bibinfo{year}{2022}).

\bibitem{rodgers2017simulation}
\bibinfo{author}{Rodgers, T.~M.}, \bibinfo{author}{Madison, J.~D.} \&
  \bibinfo{author}{Tikare, V.}
\newblock \bibinfo{title}{{Simulation of metal additive manufacturing
  microstructures using kinetic Monte Carlo}}.
\newblock \emph{\bibinfo{journal}{Computational Materials Science}}
  \textbf{\bibinfo{volume}{135}}, \bibinfo{pages}{78--89}
  (\bibinfo{year}{2017}).

\bibitem{miyoshi2017ultra}
\bibinfo{author}{Miyoshi, E.} \emph{et~al.}
\newblock \bibinfo{title}{Ultra-large-scale phase-field simulation study of
  ideal grain growth}.
\newblock \emph{\bibinfo{journal}{NPJ Computational Materials}}
  \textbf{\bibinfo{volume}{3}}, \bibinfo{pages}{1--6} (\bibinfo{year}{2017}).

\bibitem{miyoshi2021large}
\bibinfo{author}{Miyoshi, E.} \emph{et~al.}
\newblock \bibinfo{title}{{Large-scale phase-field study of anisotropic grain
  growth: Effects of misorientation-dependent grain boundary energy and
  mobility}}.
\newblock \emph{\bibinfo{journal}{Computational Materials Science}}
  \textbf{\bibinfo{volume}{186}}, \bibinfo{pages}{109992}
  (\bibinfo{year}{2021}).

\bibitem{chang2017effect}
\bibinfo{author}{Chang, K.}, \bibinfo{author}{Chen, L.-Q.},
  \bibinfo{author}{Krill~III, C.~E.} \& \bibinfo{author}{Moelans, N.}
\newblock \bibinfo{title}{{Effect of strong nonuniformity in grain boundary
  energy on 3-D grain growth behavior: A phase-field simulation study}}.
\newblock \emph{\bibinfo{journal}{Computational Materials Science}}
  \textbf{\bibinfo{volume}{127}}, \bibinfo{pages}{67--77}
  (\bibinfo{year}{2017}).

\bibitem{burke1952recrystallization}
\bibinfo{author}{Burke, J.} \& \bibinfo{author}{Turnbull, D.}
\newblock \bibinfo{title}{Recrystallization and grain growth}.
\newblock \emph{\bibinfo{journal}{Progress in metal physics}}
  \textbf{\bibinfo{volume}{3}}, \bibinfo{pages}{220--292}
  (\bibinfo{year}{1952}).

\bibitem{hillert1965theory}
\bibinfo{author}{Hillert, M.}
\newblock \bibinfo{title}{On the theory of normal and abnormal grain growth}.
\newblock \emph{\bibinfo{journal}{Acta metallurgica}}
  \textbf{\bibinfo{volume}{13}}, \bibinfo{pages}{227--238}
  (\bibinfo{year}{1965}).

\bibitem{holm2001misorientation}
\bibinfo{author}{Holm, E.~A.}, \bibinfo{author}{Hassold, G.~N.} \&
  \bibinfo{author}{Miodownik, M.~A.}
\newblock \bibinfo{title}{On misorientation distribution evolution during
  anisotropic grain growth}.
\newblock \emph{\bibinfo{journal}{Acta Materialia}}
  \textbf{\bibinfo{volume}{49}}, \bibinfo{pages}{2981--2991}
  (\bibinfo{year}{2001}).

\bibitem{qin2023grainnn}
\bibinfo{author}{Qin, Y.}, \bibinfo{author}{DeWitt, S.},
  \bibinfo{author}{Radhakrishnan, B.} \& \bibinfo{author}{Biros, G.}
\newblock \bibinfo{title}{{GrainNN: A neighbor-aware long short-term memory
  network for predicting microstructure evolution during polycrystalline grain
  formation}}.
\newblock \emph{\bibinfo{journal}{Computational Materials Science}}
  \textbf{\bibinfo{volume}{218}}, \bibinfo{pages}{111927}
  (\bibinfo{year}{2023}).

\bibitem{zollner2017topological}
\bibinfo{author}{Z{\"o}llner, D.} \& \bibinfo{author}{Rios, P.~R.}
\newblock \bibinfo{title}{{Topological changes in coarsening networks}}.
\newblock \emph{\bibinfo{journal}{Acta Materialia}}
  \textbf{\bibinfo{volume}{130}}, \bibinfo{pages}{147--154}
  (\bibinfo{year}{2017}).

\bibitem{torres2015numerical}
\bibinfo{author}{Torres, C.}, \bibinfo{author}{Emelianenko, M.},
  \bibinfo{author}{Golovaty, D.}, \bibinfo{author}{Kinderlehrer, D.} \&
  \bibinfo{author}{Ta'asan, S.}
\newblock \bibinfo{title}{{Numerical analysis of the vertex models for
  simulating grain boundary networks}}.
\newblock \emph{\bibinfo{journal}{SIAM Journal on Applied Mathematics}}
  \textbf{\bibinfo{volume}{75}}, \bibinfo{pages}{762--786}
  (\bibinfo{year}{2015}).

\bibitem{ohayon2014advanced}
\bibinfo{author}{Ohayon, R.} \& \bibinfo{author}{Soize, C.}
\newblock \emph{\bibinfo{title}{{Advanced computational vibroacoustics:
  reduced-order models and uncertainty quantification}}}
  (\bibinfo{publisher}{Cambridge University Press}, \bibinfo{year}{2014}).

\bibitem{frangos2010surrogate}
\bibinfo{author}{Frangos, M.}, \bibinfo{author}{Marzouk, Y.},
  \bibinfo{author}{Willcox, K.} \& \bibinfo{author}{van Bloemen~Waanders, B.}
\newblock \bibinfo{title}{{Surrogate and reduced-order modeling: a comparison
  of approaches for large-scale statistical inverse problems}}.
\newblock \emph{\bibinfo{journal}{Large-Scale Inverse Problems and
  Quantification of Uncertainty}} \bibinfo{pages}{123--149}
  (\bibinfo{year}{2010}).

\bibitem{legresley2000airfoil}
\bibinfo{author}{LeGresley, P.} \& \bibinfo{author}{Alonso, J.}
\newblock \bibinfo{title}{{Airfoil design optimization using reduced order
  models based on proper orthogonal decomposition}}.
\newblock In \emph{\bibinfo{booktitle}{Fluids 2000 conference and exhibit}},
  \bibinfo{pages}{2545} (\bibinfo{year}{2000}).

\bibitem{noack2011reduced}
\bibinfo{author}{Noack, B.~R.}, \bibinfo{author}{Morzynski, M.} \&
  \bibinfo{author}{Tadmor, G.}
\newblock \emph{\bibinfo{title}{{Reduced-order modelling for flow control}}},
  vol. \bibinfo{volume}{528} (\bibinfo{publisher}{Springer Science \& Business
  Media}, \bibinfo{year}{2011}).

\bibitem{li2018transfer}
\bibinfo{author}{Li, X.} \emph{et~al.}
\newblock \bibinfo{title}{A transfer learning approach for microstructure
  reconstruction and structure-property predictions}.
\newblock \emph{\bibinfo{journal}{Scientific reports}}
  \textbf{\bibinfo{volume}{8}}, \bibinfo{pages}{1--13} (\bibinfo{year}{2018}).

\bibitem{bostanabad2020reconstruction}
\bibinfo{author}{Bostanabad, R.}
\newblock \bibinfo{title}{{Reconstruction of 3D microstructures from 2D images
  via transfer learning}}.
\newblock \emph{\bibinfo{journal}{Computer-Aided Design}}
  \textbf{\bibinfo{volume}{128}}, \bibinfo{pages}{102906}
  (\bibinfo{year}{2020}).

\bibitem{goodfellow2014generative}
\bibinfo{author}{Goodfellow, I.} \emph{et~al.}
\newblock \bibinfo{title}{{Generative adversarial nets}}.
\newblock \emph{\bibinfo{journal}{Advances in neural information processing
  systems}} \textbf{\bibinfo{volume}{27}} (\bibinfo{year}{2014}).

\bibitem{yang2018microstructural}
\bibinfo{author}{Yang, Z.} \emph{et~al.}
\newblock \bibinfo{title}{{Microstructural materials design via deep
  adversarial learning methodology}}.
\newblock \emph{\bibinfo{journal}{Journal of Mechanical Design}}
  \textbf{\bibinfo{volume}{140}} (\bibinfo{year}{2018}).

\bibitem{lee2021fast}
\bibinfo{author}{Lee, X.~Y.} \emph{et~al.}
\newblock \bibinfo{title}{{Fast inverse design of microstructures via
  generative invariance networks}}.
\newblock \emph{\bibinfo{journal}{Nature Computational Science}}
  \textbf{\bibinfo{volume}{1}}, \bibinfo{pages}{229--238}
  (\bibinfo{year}{2021}).

\bibitem{hochreiter1997long}
\bibinfo{author}{Hochreiter, S.} \& \bibinfo{author}{Schmidhuber, J.}
\newblock \bibinfo{title}{Long short-term memory}.
\newblock \emph{\bibinfo{journal}{Neural computation}}
  \textbf{\bibinfo{volume}{9}}, \bibinfo{pages}{1735--1780}
  (\bibinfo{year}{1997}).

\bibitem{Montes-de-Oca-Zapiain:2021uj}
\bibinfo{author}{Montes~de Oca~Zapiain, D.}, \bibinfo{author}{Stewart, J.~A.}
  \& \bibinfo{author}{Dingreville, R.}
\newblock \bibinfo{title}{Accelerating phase-field-based microstructure
  evolution predictions via surrogate models trained by machine learning
  methods}.
\newblock \emph{\bibinfo{journal}{npj Computational Materials}}
  \textbf{\bibinfo{volume}{7}}, \bibinfo{pages}{3} (\bibinfo{year}{2021}).

\bibitem{hu2022accelerating}
\bibinfo{author}{Hu, C.}, \bibinfo{author}{Martin, S.} \&
  \bibinfo{author}{Dingreville, R.}
\newblock \bibinfo{title}{Accelerating phase-field predictions via recurrent
  neural networks learning the microstructure evolution in latent space}.
\newblock \emph{\bibinfo{journal}{Computer Methods in Applied Mechanics and
  Engineering}} \textbf{\bibinfo{volume}{397}}, \bibinfo{pages}{115128}
  (\bibinfo{year}{2022}).

\bibitem{yang2021self}
\bibinfo{author}{Yang, K.} \emph{et~al.}
\newblock \bibinfo{title}{{Self-supervised learning and prediction of
  microstructure evolution with convolutional recurrent neural networks}}.
\newblock \emph{\bibinfo{journal}{Patterns}} \textbf{\bibinfo{volume}{2}},
  \bibinfo{pages}{100243} (\bibinfo{year}{2021}).

\bibitem{oommen2022learning}
\bibinfo{author}{Oommen, V.}, \bibinfo{author}{Shukla, K.},
  \bibinfo{author}{Goswami, S.}, \bibinfo{author}{Dingreville, R.} \&
  \bibinfo{author}{Karniadakis, G.~E.}
\newblock \bibinfo{title}{Learning two-phase microstructure evolution using
  neural operators and autoencoder architectures}.
\newblock \emph{\bibinfo{journal}{npj Computational Materials}}
  \textbf{\bibinfo{volume}{8}}, \bibinfo{pages}{190} (\bibinfo{year}{2022}).

\bibitem{lu2021learning}
\bibinfo{author}{Lu, L.}, \bibinfo{author}{Jin, P.}, \bibinfo{author}{Pang,
  G.}, \bibinfo{author}{Zhang, Z.} \& \bibinfo{author}{Karniadakis, G.~E.}
\newblock \bibinfo{title}{Learning nonlinear operators via deeponet based on
  the universal approximation theorem of operators}.
\newblock \emph{\bibinfo{journal}{Nature machine intelligence}}
  \textbf{\bibinfo{volume}{3}}, \bibinfo{pages}{218--229}
  (\bibinfo{year}{2021}).

\bibitem{syha2009generalized}
\bibinfo{author}{Syha, M.} \& \bibinfo{author}{Weygand, D.}
\newblock \bibinfo{title}{A generalized vertex dynamics model for grain growth
  in three dimensions}.
\newblock \emph{\bibinfo{journal}{Modelling and Simulation in Materials Science
  and Engineering}} \textbf{\bibinfo{volume}{18}}, \bibinfo{pages}{015010}
  (\bibinfo{year}{2009}).

\bibitem{kawasaki1989vertex}
\bibinfo{author}{Kawasaki, K.}, \bibinfo{author}{Nagai, T.} \&
  \bibinfo{author}{Nakashima, K.}
\newblock \bibinfo{title}{Vertex models for two-dimensional grain growth}.
\newblock \emph{\bibinfo{journal}{Philosophical Magazine B}}
  \textbf{\bibinfo{volume}{60}}, \bibinfo{pages}{399--421}
  (\bibinfo{year}{1989}).

\bibitem{wakai2000three}
\bibinfo{author}{Wakai, F.}, \bibinfo{author}{Enomoto, N.} \&
  \bibinfo{author}{Ogawa, H.}
\newblock \bibinfo{title}{Three-dimensional microstructural evolution in ideal
  grain growth—general statistics}.
\newblock \emph{\bibinfo{journal}{Acta Materialia}}
  \textbf{\bibinfo{volume}{48}}, \bibinfo{pages}{1297--1311}
  (\bibinfo{year}{2000}).

\bibitem{xue2022physics}
\bibinfo{author}{Xue, T.}, \bibinfo{author}{Gan, Z.}, \bibinfo{author}{Liao,
  S.} \& \bibinfo{author}{Cao, J.}
\newblock \bibinfo{title}{Physics-embedded graph network for accelerating
  phase-field simulation of microstructure evolution in additive
  manufacturing}.
\newblock \emph{\bibinfo{journal}{npj Computational Materials}}
  \textbf{\bibinfo{volume}{8}}, \bibinfo{pages}{201} (\bibinfo{year}{2022}).

\bibitem{dai2021graph}
\bibinfo{author}{Dai, M.}, \bibinfo{author}{Demirel, M.~F.},
  \bibinfo{author}{Liang, Y.} \& \bibinfo{author}{Hu, J.-M.}
\newblock \bibinfo{title}{{Graph neural networks for an accurate and
  interpretable prediction of the properties of polycrystalline materials}}.
\newblock \emph{\bibinfo{journal}{npj Computational Materials}}
  \textbf{\bibinfo{volume}{7}}, \bibinfo{pages}{103} (\bibinfo{year}{2021}).

\bibitem{hestroffer2023graph}
\bibinfo{author}{Hestroffer, J.~M.}, \bibinfo{author}{Charpagne, M.-A.},
  \bibinfo{author}{Latypov, M.~I.} \& \bibinfo{author}{Beyerlein, I.~J.}
\newblock \bibinfo{title}{{Graph neural networks for efficient learning of
  mechanical properties of polycrystals}}.
\newblock \emph{\bibinfo{journal}{Computational Materials Science}}
  \textbf{\bibinfo{volume}{217}}, \bibinfo{pages}{111894}
  (\bibinfo{year}{2023}).

\bibitem{kipf2016semi}
\bibinfo{author}{Kipf, T.~N.} \& \bibinfo{author}{Welling, M.}
\newblock \bibinfo{title}{{Semi-supervised classification with graph
  convolutional networks}}.
\newblock \emph{\bibinfo{journal}{arXiv preprint arXiv:1609.02907}}
  (\bibinfo{year}{2016}).

\bibitem{scarselli2008graph}
\bibinfo{author}{Scarselli, F.}, \bibinfo{author}{Gori, M.},
  \bibinfo{author}{Tsoi, A.~C.}, \bibinfo{author}{Hagenbuchner, M.} \&
  \bibinfo{author}{Monfardini, G.}
\newblock \bibinfo{title}{The graph neural network model}.
\newblock \emph{\bibinfo{journal}{IEEE transactions on neural networks}}
  \textbf{\bibinfo{volume}{20}}, \bibinfo{pages}{61--80}
  (\bibinfo{year}{2008}).

\bibitem{morris2019weisfeiler}
\bibinfo{author}{Morris, C.} \emph{et~al.}
\newblock \bibinfo{title}{{Weisfeiler and leman go neural: Higher-order graph
  neural networks}}.
\newblock In \emph{\bibinfo{booktitle}{Proceedings of the AAAI conference on
  artificial intelligence}}, vol.~\bibinfo{volume}{33},
  \bibinfo{pages}{4602--4609} (\bibinfo{year}{2019}).

\bibitem{wieder2020compact}
\bibinfo{author}{Wieder, O.} \emph{et~al.}
\newblock \bibinfo{title}{A compact review of molecular property prediction
  with graph neural networks}.
\newblock \emph{\bibinfo{journal}{Drug Discovery Today: Technologies}}
  \textbf{\bibinfo{volume}{37}}, \bibinfo{pages}{1--12} (\bibinfo{year}{2020}).

\bibitem{duvenaud2015convolutional}
\bibinfo{author}{Duvenaud, D.~K.} \emph{et~al.}
\newblock \bibinfo{title}{Convolutional networks on graphs for learning
  molecular fingerprints}.
\newblock \emph{\bibinfo{journal}{Advances in neural information processing
  systems}} \textbf{\bibinfo{volume}{28}} (\bibinfo{year}{2015}).

\bibitem{gilmer2017neural}
\bibinfo{author}{Gilmer, J.}, \bibinfo{author}{Schoenholz, S.~S.},
  \bibinfo{author}{Riley, P.~F.}, \bibinfo{author}{Vinyals, O.} \&
  \bibinfo{author}{Dahl, G.~E.}
\newblock \bibinfo{title}{{Neural message passing for quantum chemistry}}.
\newblock In \emph{\bibinfo{booktitle}{International conference on machine
  learning}}, \bibinfo{pages}{1263--1272} (\bibinfo{organization}{PMLR},
  \bibinfo{year}{2017}).

\bibitem{xie2018crystal}
\bibinfo{author}{Xie, T.} \& \bibinfo{author}{Grossman, J.~C.}
\newblock \bibinfo{title}{{Crystal graph convolutional neural networks for an
  accurate and interpretable prediction of material properties}}.
\newblock \emph{\bibinfo{journal}{Physical review letters}}
  \textbf{\bibinfo{volume}{120}}, \bibinfo{pages}{145301}
  (\bibinfo{year}{2018}).

\bibitem{chen2022gc}
\bibinfo{author}{Chen, J.}, \bibinfo{author}{Wang, X.} \& \bibinfo{author}{Xu,
  X.}
\newblock \bibinfo{title}{{GC-LSTM: Graph convolution embedded LSTM for dynamic
  network link prediction}}.
\newblock \emph{\bibinfo{journal}{Applied Intelligence}} \bibinfo{pages}{1--16}
  (\bibinfo{year}{2022}).

\bibitem{pareja2020evolvegcn}
\bibinfo{author}{Pareja, A.} \emph{et~al.}
\newblock \bibinfo{title}{{Evolvegcn: Evolving graph convolutional networks for
  dynamic graphs}}.
\newblock In \emph{\bibinfo{booktitle}{Proceedings of the AAAI conference on
  artificial intelligence}}, vol.~\bibinfo{volume}{34},
  \bibinfo{pages}{5363--5370} (\bibinfo{year}{2020}).

\bibitem{goyal2020dyngraph2vec}
\bibinfo{author}{Goyal, P.}, \bibinfo{author}{Chhetri, S.~R.} \&
  \bibinfo{author}{Canedo, A.}
\newblock \bibinfo{title}{{dyngraph2vec: Capturing network dynamics using
  dynamic graph representation learning}}.
\newblock \emph{\bibinfo{journal}{Knowledge-Based Systems}}
  \textbf{\bibinfo{volume}{187}}, \bibinfo{pages}{104816}
  (\bibinfo{year}{2020}).

\bibitem{vaswani2017attention}
\bibinfo{author}{Vaswani, A.} \emph{et~al.}
\newblock \bibinfo{title}{Attention is all you need}.
\newblock In \emph{\bibinfo{booktitle}{Advances in neural information
  processing systems}}, \bibinfo{pages}{5998--6008} (\bibinfo{year}{2017}).

\bibitem{pinomaa2019process}
\bibinfo{author}{Pinomaa, T.} \emph{et~al.}
\newblock \bibinfo{title}{{Process-Structure-Properties-Performance modeling
  for selective laser melting}}.
\newblock \emph{\bibinfo{journal}{Metals}} \textbf{\bibinfo{volume}{9}},
  \bibinfo{pages}{1138} (\bibinfo{year}{2019}).

\bibitem{qin2022dendrite}
\bibinfo{author}{Qin, Y.}, \bibinfo{author}{Bao, Y.}, \bibinfo{author}{DeWitt,
  S.}, \bibinfo{author}{Radhakrishnan, B.} \& \bibinfo{author}{Biros, G.}
\newblock \bibinfo{title}{Dendrite-resolved, full-melt-pool phase-field
  simulations to reveal non-steady-state effects and to test an approximate
  model}.
\newblock \emph{\bibinfo{journal}{Computational Materials Science}}
  \textbf{\bibinfo{volume}{207}}, \bibinfo{pages}{111262}
  (\bibinfo{year}{2022}).

\bibitem{Tourret2015a}
\bibinfo{author}{Tourret, D.} \& \bibinfo{author}{Karma, A.}
\newblock \bibinfo{title}{{Growth competition of columnar dendritic grains: A
  phase-field study}}.
\newblock \emph{\bibinfo{journal}{Acta Materialia}}
  \textbf{\bibinfo{volume}{82}}, \bibinfo{pages}{64--83}
  (\bibinfo{year}{2015}).

\bibitem{takaki2018competitive}
\bibinfo{author}{Takaki, T.} \emph{et~al.}
\newblock \bibinfo{title}{{Competitive grain growth during directional
  solidification of a polycrystalline binary alloy: Three-dimensional
  large-scale phase-field study}}.
\newblock \emph{\bibinfo{journal}{Materialia}} \textbf{\bibinfo{volume}{1}},
  \bibinfo{pages}{104--113} (\bibinfo{year}{2018}).

\bibitem{bragard2002linking}
\bibinfo{author}{Bragard, J.}, \bibinfo{author}{Karma, A.},
  \bibinfo{author}{Lee, Y.~H.} \& \bibinfo{author}{Plapp, M.}
\newblock \bibinfo{title}{Linking phase-field and atomistic simulations to
  model dendritic solidification in highly undercooled melts}.
\newblock \emph{\bibinfo{journal}{Interface Science}}
  \textbf{\bibinfo{volume}{10}}, \bibinfo{pages}{121--136}
  (\bibinfo{year}{2002}).

\bibitem{shi2020masked}
\bibinfo{author}{Shi, Y.} \emph{et~al.}
\newblock \bibinfo{title}{{Masked label prediction: Unified message passing
  model for semi-supervised classification}}.
\newblock \emph{\bibinfo{journal}{arXiv preprint arXiv:2009.03509}}
  (\bibinfo{year}{2020}).

\bibitem{vedantam2006efficient}
\bibinfo{author}{Vedantam, S.} \& \bibinfo{author}{Patnaik, B.}
\newblock \bibinfo{title}{Efficient numerical algorithm for multiphase field
  simulations}.
\newblock \emph{\bibinfo{journal}{Physical Review E}}
  \textbf{\bibinfo{volume}{73}}, \bibinfo{pages}{016703}
  (\bibinfo{year}{2006}).

\bibitem{Badillo2006}
\bibinfo{author}{Badillo, A.} \& \bibinfo{author}{Beckermann, C.}
\newblock \bibinfo{title}{{Phase-field simulation of the columnar-to-equiaxed
  transition in alloy solidification}}.
\newblock \emph{\bibinfo{journal}{Acta Materialia}}
  \textbf{\bibinfo{volume}{54}}, \bibinfo{pages}{2015--2026}
  (\bibinfo{year}{2006}).

\bibitem{Pinomaa2020}
\bibinfo{author}{Pinomaa, T.} \emph{et~al.}
\newblock \bibinfo{title}{{Phase field modeling of rapid resolidification of
  Al-Cu thin films}}.
\newblock \emph{\bibinfo{journal}{Journal of Crystal Growth}}
  \textbf{\bibinfo{volume}{532}} (\bibinfo{year}{2020}).

\bibitem{xu2022three}
\bibinfo{author}{Xu, F.}, \bibinfo{author}{Xiong, F.}, \bibinfo{author}{Li,
  M.-J.} \& \bibinfo{author}{Lian, Y.}
\newblock \bibinfo{title}{{Three-Dimensional Numerical Simulation of Grain
  Growth during Selective Laser Melting of 316L Stainless Steel}}.
\newblock \emph{\bibinfo{journal}{Materials}} \textbf{\bibinfo{volume}{15}},
  \bibinfo{pages}{6800} (\bibinfo{year}{2022}).

\bibitem{zhang2018improved}
\bibinfo{author}{Zhang, Z.}
\newblock \bibinfo{title}{Improved adam optimizer for deep neural networks}.
\newblock In \emph{\bibinfo{booktitle}{2018 IEEE/ACM 26th International
  Symposium on Quality of Service (IWQoS)}}, \bibinfo{pages}{1--2}
  (\bibinfo{organization}{Ieee}, \bibinfo{year}{2018}).

\bibitem{lookman2019active}
\bibinfo{author}{Lookman, T.}, \bibinfo{author}{Balachandran, P.~V.},
  \bibinfo{author}{Xue, D.} \& \bibinfo{author}{Yuan, R.}
\newblock \bibinfo{title}{Active learning in materials science with emphasis on
  adaptive sampling using uncertainties for targeted design}.
\newblock \emph{\bibinfo{journal}{npj Computational Materials}}
  \textbf{\bibinfo{volume}{5}}, \bibinfo{pages}{1--17} (\bibinfo{year}{2019}).

\bibitem{krogh1994neural}
\bibinfo{author}{Krogh, A.} \& \bibinfo{author}{Vedelsby, J.}
\newblock \bibinfo{title}{Neural network ensembles, cross validation, and
  active learning}.
\newblock \emph{\bibinfo{journal}{Advances in neural information processing
  systems}} \textbf{\bibinfo{volume}{7}} (\bibinfo{year}{1994}).

\end{thebibliography}
